\def\jcap{JCAP}

\def \<{\langle}
\def \>{\rangle}

\newcommand{\ra}{\;\raise1.0pt\hbox{$'$}\hskip-6pt\partial\;}
\newcommand{\lo}{\;\overline{\raise1.0pt\hbox{$'$}\hskip-6pt\partial}\;}

\newcommand{\Abs}{\abstract}

\documentclass[a4paper,11pt]{article}
\pdfoutput=1

\usepackage{jcappub,graphicx,epsfig,natbib,color,times,bm,amsmath,multirow,hyperref}
\usepackage[ddmmyyyy,hhmmss]{datetime}
\usepackage[utf8]{inputenc}
\usepackage{physics}
\usepackage{fullpage}
\usepackage{graphicx}

\title{Noise residuals for GW150914 using maximum likelihood and numerical
relativity templates }

\author[a]{Andrew {D. Jackson},}\emailAdd{jackson@nbi.dk}

\author[b,c]{Hao Liu,}\emailAdd{liuhao@nbi.dk}

\author[b]{and Pavel Naselsky}\emailAdd{naselsky@nbi.dk}

\affiliation[a]{The Niels Bohr International Academy, Blegdamsvej 17, DK-2100
Copenhagen, Denmark}

\affiliation[b]{The Niels Bohr Institute \& Discovery Center, Blegdamsvej 17,
DK-2100 Copenhagen, Denmark}

\affiliation[c]{Key Laboratory of Particle and Astrophysics, Institute of High
Energy Physics, CAS, 19B YuQuan Road, Beijing, China}

\begin{document}

\Abs{

We reexamine the results presented in~\cite{2019JCAP...02..019N} in which the
properties of the noise residuals in the 40\,ms chirp domain of GW150914 were
investigated. This paper confirmed the presence of strong (i.e., about 0.80)
correlations between residual noise in the Hanford and Livingston detectors in
the chirp domain as previously seen by~\cite{2017JCAP...08..013C} when using a
numerical relativity template given in~\cite{PhysRevLett.116.061102}. It was
also shown in~\cite{2019JCAP...02..019N} that a so-called maximum likelihood
template can reduce these statistically significant cross-correlations. Here,
we demonstrate that the reduction of correlation and statistical significance
is due to (i) the use of a peculiar template with extreme spin (0.977), which
is qualitatively different from the properties of GW150914 originally
published by LIGO, (ii) a suspicious MCMC chain, (iii) uncertainties in the
matching of the maximum likelihood (ML) template to the data in the Fourier
domain, and (iv) a biased estimation of the significance that gives
counterintuitive results. We show that rematching the maximum likelihood
template to the data in the 0.2\,s domain containing the GW150914 signal
restores these correlations at the level of $60\%$ of those found
previously~\cite{2019JCAP...02..019N}. With necessary corrections, the
probability given in~\cite{2019JCAP...02..019N} for the residual correlation
will decrease by more than one order of magnitude. Since the ML template is
itself problematic, results associated with it are illustrative rather than
final.

}

\maketitle

\section{Introduction and Motivation.}\label{sec:intro}

Knowledge of the properties of noise in the LIGO Hanford and Livingston data
sets is crucial for verification of the methods used to determined the
physical nature and quantitative details of the GW150914 event and all
subsequent events~\cite{PhysRevLett.116.061102}. In practice, the maximum
likelihood method used in \cite{2019JCAP...02..019N} to analyze the GW events
effectively assumes that the noise is both stationary and Gaussian. The
validity of these crucial assumptions can be tested by considering the
cross-correlations of the Hanford and Livingston noise residuals, which should
be at the level of chance correlations. In~\cite{2017JCAP...08..013C} we
explored the assumption of uncorrelated Hanford/Livingston noise residuals for
GW150914 by considering a windowed Pearson cross-correlator (see
Appendix~\ref{app:correlator} for more details). Significant abnormal
correlations were found.

Using a numerical relativity (NR) template~\cite{PhysRevLett.116.061102}, we
identified~\cite{2017JCAP...08..013C} three 40\,ms domains of significant
noise correlation in a 0.2\,s record\footnote{The strain data for a 0.2\,s
time interval including GW150914 and the NR template are available at the LIGO
Open Science Center (LOSC). In the present paper, time positions are measured
relative to the 32 seconds centered at GPS time 1126259462, a convention
established in some of the files of the initial LOSC release. For example, the
peak amplitude of the GW150914 event occurs at approximately 16.42 seconds.}:
the chirp domain (16.39-16.43s) with a Hanford/Livingston cross-correlation of
$0.80$, the precursor domain (16.27-16.31s) with a cross-correlation of
$0.60$, and the echo domain (16.47-16.51s) during ring-down with a
cross-correlation of similar amplitude~\cite{2017JCAP...08..013C}.  Taken
together, these results are not consistent with the assumption of uncorrelated
noise residuals under Gaussian and stationary assumptions.  Recently, this
result was confirmed using the same NR template and $0.2$s time interval
in~\cite{2019JCAP...02..019N}.  However, the authors of this paper also claim
that the strong abnormal residual correlations discovered
in~\cite{2017JCAP...08..013C} can be reduced significantly in the chirp domain
by using an alternative ML template. We cannot with certainty exclude the
possibility that some template in LIGO's models of binary black hole mergers
can locally reduce residual correlations.  While we investigated this question
in~\cite{2019JCAP...02..019N}, we did not consider specific templates
involving black holes with extreme spins. Given that in the ML template, one
of its participating black hole has a spin indistinguishably close to the
maximum possible value of 1, the ML template adopted
in~\cite{2019JCAP...02..019N} is an example of one such extreme template.

Clearly, independent verification of the methods and results of LIGO, such
as~\cite{2019JCAP...02..019N}, is a matter of importance for our understanding
of black hole physics. Especially since, according to ref.~[37]
in~\cite{PhysRevLett.116.061102}, the NR template used by LIGO is SXS:BBH:0305
which has on-axis spins of $+0.33$ and $-0.44$ is almost identical to a zero
spin template. In sharp contrast, the ML template discussed
in~\cite{2019JCAP...02..019N} involves a black hole with extreme spin ---
almost the physical limit. This difference has important consequences. Unlike
low-spin black holes, high-spin Kerr black holes cannot be formed at early
stages in the evolution of the Universe as so-called primordial black holes.
The black holes suggested by the results of~\cite{2019JCAP...02..019N} would
rather indicate that the black holes involved have a stellar origin from very
massive stars.  Even given the stellar origin of such black holes, the extreme
spin would appear to be challenging.

We also point out that the cleaned (i.e., bandpassed and notch-filtered)
strain data used by~\cite{2019JCAP...02..019N} is essentially identical to
ours~\cite{2017JCAP...08..013C, NBI:Gravitational.waves}. Using a bandpass
window of $35\le f\le 350$\,Hz, none of the 37240 templates considered
in~\cite{2018arXiv180710312B} (from which the ML template is extracted) can
reduce the precursor effect observed during the first $0.1$s of GW150914.
(This leads to a cross-correlation of $0.60$ between H/L residuals.)  Indeed,
in this time window the NR template, like all other templates filtered from
$35\le f\le 350$\,Hz, is almost negligible in comparison with noise.

Our goal here is to reevaluate the methods and results
of~\cite{2019JCAP...02..019N} with regard to their claims about the noise
residuals for GW150914. We will first show that the matching of the ML
template to the data depends on the estimator assumed.
In~\cite{2019JCAP...02..019N} the quantity to be optimized is a complex
overlap function (of Hanford and Livingston strain data) with the same overall
phase for each template.  (See Table~\ref{tab:alex params} and its caption.)
This constraint on the phase proves to be crucial for introducing an effective
shift of the peak position in the chirp domain. Consequently, it plays a major
role in decreasing the residual noise correlation.  Second, calculation of the
SNR parameter involves a power spectral density (PSD) obtained from a time
interval that is nearly 4 orders of magnitude larger than the length of the
most significant GW domain (1040\,s versus 0.2\,s). The validity of this
approach critically depends on the assumptions that the noise is stationary
and Gaussian. Neither assumption is true.  On the other hand, the equal-time
comparison used for matching with the Pearson estimator of eq.~(\ref{corr}) does
not depend on any such assumptions. It can be applied safely for short (e.g.,
0.2\,s) records.  In general, it is important to bear in mind that the
implementation of any statistical concepts, such as chance correlations, for
the estimation of the significance of noise cross-correlations needs to be
performed with care.  Due to the non-stationarity and non-Gaussianity of the
noise residuals, all such estimates can be misleading. (This point will be
illustrated in Sections 4-5.) We emphasize that we do not regard such
assumptions as safe and have avoided them in~\cite{,2019JCAP...02..019N}
and~\cite{2017JCAP...08..013C}.  Nevertheless, we will from time to time be
forced to assume stationarity and Gaussianity in the interests of following
the methods of~\cite{2019JCAP...02..019N}.

The outline of the paper is as follows.  In Section~\ref{sec:ML brief}, we
briefly review the maximum likelihood approach and the corresponding templates
as provided in~\cite{2018arXiv180710312B} and used
in~\cite{2019JCAP...02..019N}.  In Section~\ref{sec:HL residual and cc}, we
consider the residuals and their correlations for both this template and for
the NR template. The significance of the residual correlations is estimated in
Section~\ref{sec:resi-CC significance}.  Since we regard the ML template
itself as unsatisfactory, our primary focus is on how to
correct~\cite{2019JCAP...02..019N} in order to obtain unbiased estimations.
We investigate some uncertainties in the ML method in
Section~\ref{sec:uncertain}, and a brief discussion of the results is given in
Section~\ref{sec:conclusion}.

While this paper is largely concerned with residual correlations, the
following should be kept in mind: The presence of statistically significant
correlations in the Hanford and Livingston residuals is sufficient to
eliminate any proposed waveform (e.g., the NR template
of~\cite{PhysRevLett.116.061102}) from further consideration.  The absence of
such residual correlations does {\em not\/} provide any evidence that the
proposed waveform is correct.

\section{The ``maximum likelihood'' template}\label{sec:ML brief}

\subsection{Parameters of the ML-template}\label{sub:ML-params}

In Table~\ref{tab:alex params}, we summarize the parameters of the ML template
used by~\cite{2019JCAP...02..019N}, which was produced
in~\cite{2018arXiv180710312B}.
\begin{table}[!htb]
 \centering
 \begin{tabular}{|r|l|r|r|} \hline
    No. & Name                              &   Old value   & New value   \\ \hline
    1   & ra ($\alpha$)                     &   1.5730257   &   2.1140412 \\ \hline
    2   & dec ($\delta$)                    &  -1.2734810   &  -1.2518563 \\ \hline
    3   & distance   ($d_L$)                & 476.7564547   & 527.0627598 \\ \hline
    4   & inclination ($\iota$)             &   2.9132713   &   3.0021228 \\ \hline
    5   & mass1 ($m_1$)                     &  39.0257656   &  38.4900335 \\ \hline
    6   & mass2 ($m_2$)                     &  32.0625631   &  32.3104771 \\ \hline
    7   & polarization  ($\psi$)            &   5.9925231   &   2.6053099 \\ \hline
    8   & spin1\_a      ($a_1$)             &   0.9767961   &   0.9635978 \\ \hline
    9   & spin1\_azimuthal ($\theta_1^a$)   &   3.6036952   &   4.7164805 \\ \hline
    10  & spin1\_polar  ($\theta_1^p$)      &   1.6283548   &   1.9250337 \\ \hline
    11  & spin2\_a  ($a_2$)                 &   0.1887608   &   0.2894704 \\ \hline
    12  & spin2\_azimuthal ($\theta_2^a)$   &   3.4359460   &   2.0230135 \\ \hline
    13  & spin2\_polar ($\theta_2^p)$       &   2.4915268   &   0.7928019 \\ \hline
    14  & tc (from 1126259462)              &   0.4175646   &   0.4151170 \\ \hline\hline
    15a & coa\_phase                        &   0.6883212   &    N/A      \\ \hline
    15b & phase\_shift $\phi_0$             &  -0.9155276   &   1.7576289 \\ \hline
    15c & summed phase                      &  -0.2272064   &   1.7576289 \\ \hline\hline
        & SNR                               &  24.36169     &  24.33653   \\ \hline
 \end{tabular}
 \caption{The parameters of the ML template for GW150914 used
    by~\cite{2019JCAP...02..019N}. The large number of digits is especially
    important for ``tc'' in order to ensure that the uncertainty in
    positioning is much less than the possible $\pm 10$\,ms delay in the
    arrival time.  The ML parameters (in the column ``Old values'') were taken
    from~\cite{2018arXiv180710312B}. During the preparation of this work,
    however, all of these parameters were updated
    online~\cite{gw150914:mcmc:ml:online}, but~\cite{2019JCAP...02..019N} used
    only the old values. Nevertheless, we append the new values for this table
    which involve relative changes in the parameters ranging from $2\%$ to
    $200\%$.  In spite of these changes, the resulting templates are almost
    identical. In~\cite{2019JCAP...02..019N}, the final parameter (15a) is
    unused in the Markov Chain Monte Carlo (MCMC) process. Its value must be
    combined with the externally determined $\phi_0$ (15b) to give the summed
    phase (15c).  The respective SNRs are shown in the final row. Although
    they are not significantly different, the original template has a slightly
    higher SNR. We prefer the old template in the following analysis, not only
    because of the higher SNR but also because this is precisely the template
    used by~\cite{2019JCAP...02..019N}. }
 \label{tab:alex params}
\end{table}

Table~\ref{tab:alex params} is characterized by 14 parameters that are
actively involved in the MCMC optimization and a single parameter (either
${\rm coa\_phase}$ or $\phi_0$) that is determined a posteriori.  During the
random walk, $\phi_0$ is a meaningless random number. When the MCMC process is
finished, a new estimation will be given as $\phi_0$ so that the final
template is
\begin{equation}
    \tilde{h}(\theta', \phi_0) = \tilde{h}^0(\theta') e^{i \phi_0} \ .
\label{eq4}
\end{equation}
Here, $h^0$ is the template with $\phi_0 = 0$, and $\theta'$
represents the 15 parameters including random $\phi_0$. 

\begin{figure}
    \centerline{
    \includegraphics[width=0.5\textwidth]{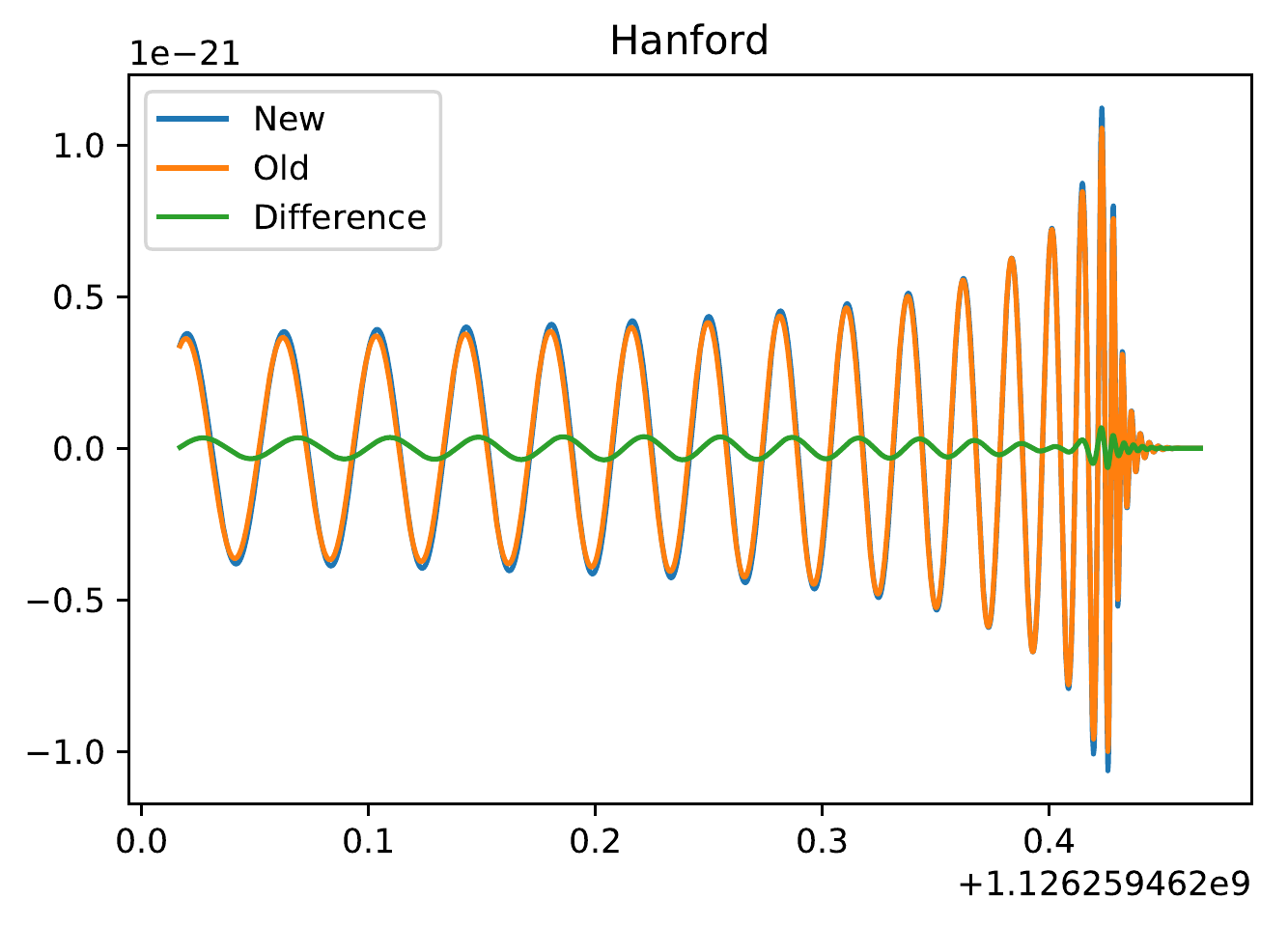}
    \includegraphics[width=0.5\textwidth]{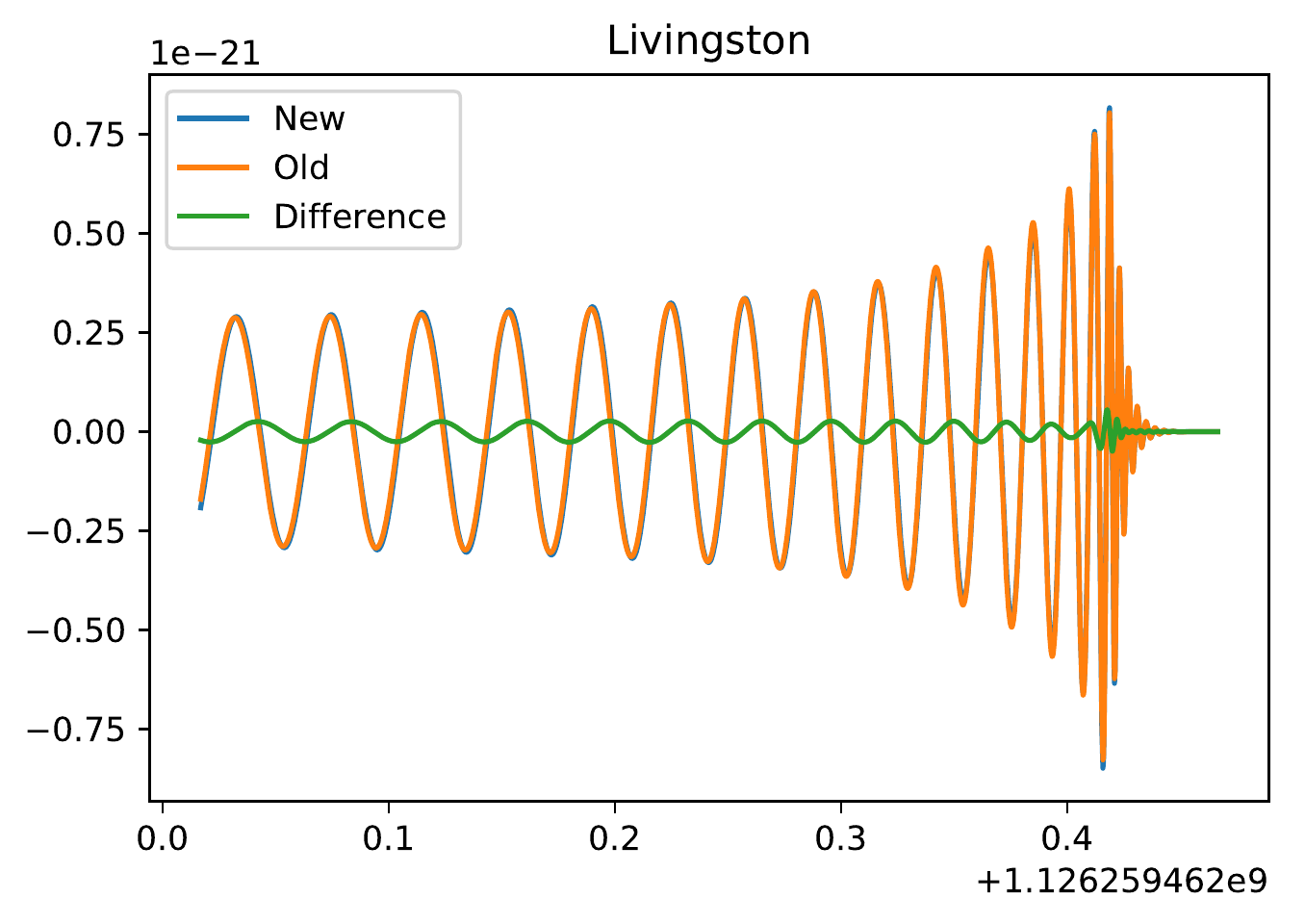}
    }
    \caption{Left panel: Comparison of the old and new templates from
      Table~\ref{tab:alex params} for Hanford. Right panel: The same for
      Livingston.The green line is the difference between the old and new
      templates. }
    \label{fa1}
\end{figure}

Note that, during the preparation of this work, all of the MCMC parameters
were updated online~\cite{gw150914:mcmc:ml:online},
but~\cite{2019JCAP...02..019N} still uses the old values. In Figure~\ref{fa1}
we show the old and new templates from Table~\ref{tab:alex params}. Although
the variations in the parameters themselves range from $2$ to $200\%$ and the
change in the summed phase $\phi_0$ is large, the differences in the time
domain morphology are quite small. Moreover, the two templates have almost
identical SNRs with a relative variation $0.1\%$, which is negligible given
the magnitude of the noise. This is a clear reflection of strong degeneracy in
the parameters\footnote{These degeneracies have been discussed
in~\cite{Degeneracy}. See, for example, the left panel of Figure 7 there.}.
Note in particular that the 2.4\,ms in the time shift, tc, is actually
remarkably large.  (In this regard, see Figure~\ref{fa1} below.) These
differences between the old and new parameters immediately tell us that an
expensive ML approach has little scientific merit in determining the possible
nature of black hole binaries in events like GW150914. Although the entire ML
approach employed in~\cite{2018arXiv180710312B} is of questionable value, we
will simply take the ML template used by~\cite{2019JCAP...02..019N} for some
further analysis.

Normally, for an MCMC chain, the ML point is asymptotically a best fit point,
whereas other points are random walks and can not be regarded as properly
fitted. In fact, even the ML point is not a genuine best fit point, because
when the MCMC approach is re-run, a new ML point will be generated, which is
normally \emph{not} the previous one. This is excellently illustrated by the
two columns in Table~\ref{tab:alex params} for the new and old parameters:
each of them is the ML point for its own round, but they give significantly
different parameters. On the other hand, a fitting procedure, if possible, can
give the unique best fit. In the case of two detectors like GW150914, the
projection parameters (rows 1, 2, 3, 4, 7, 14, 15 in Table~\ref{tab:alex
params}) can be related to the matching (fitting) parameters (amplitudes,
phases, arrival times) by a physical connection, as shown
by~\cite{2007PhDT.......290B}. For two detectors, the number of degrees of
freedom for matching is less than the number of physical parameter, a
re-matching procedure can be applied to the points in the MCMC chain to
improve the matching with data. This will be done below.

\subsection{Self-contradictions in the MCMC chain}\label{sub:contradiction}

\begin{figure}
\begin{center}
    \includegraphics[width=0.96\textwidth]{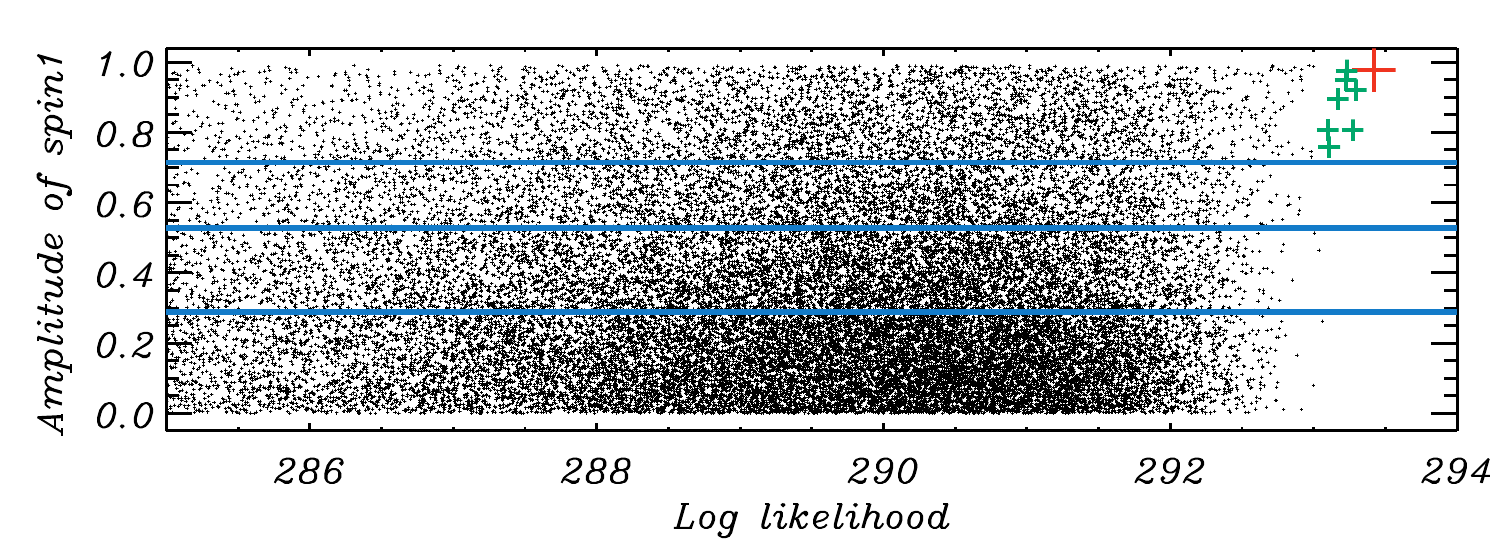}
    \includegraphics[width=0.96\textwidth]{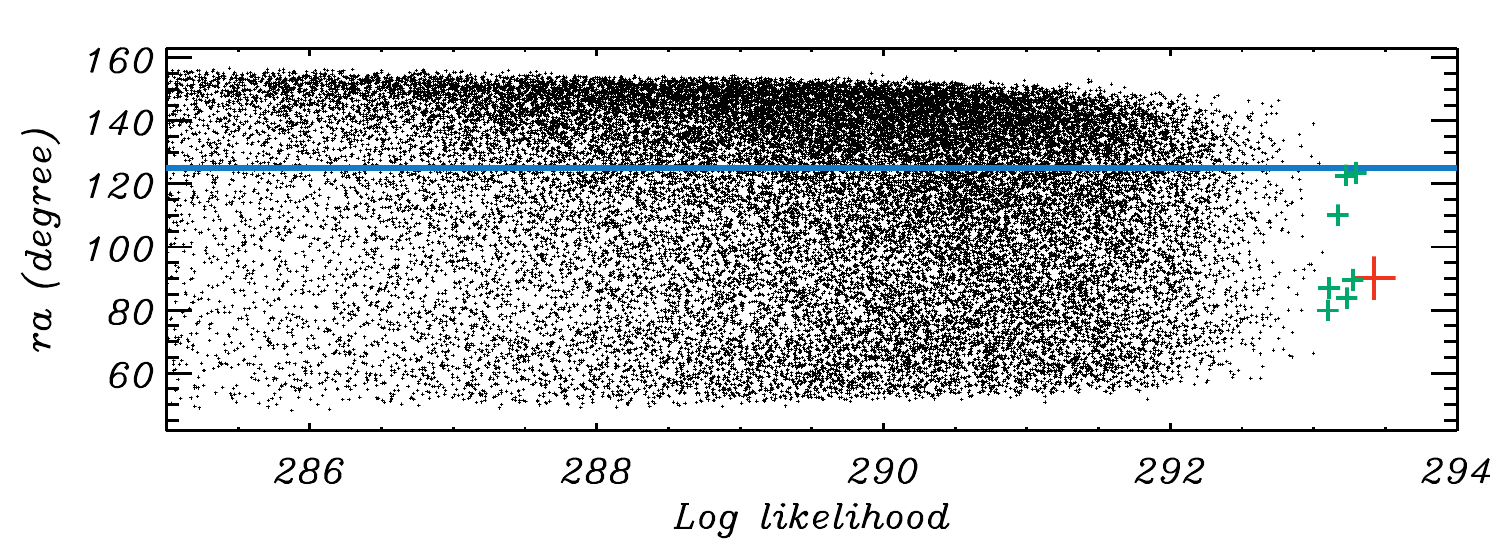}
\end{center}
    \caption{\emph{Upper}: Scatter plot of the spin1 amplitude versus the
    log-likelihood. The blue lines are 1/2 (median), 1/4 and 1/8 splits of all
    templates. \emph{Lower}: the same for ra, with the blue line for median
    split. Red cross for the ML template, and green crosses for the next 7
    highest likelihood templates. Data from the MCMC chain used
    by~\cite{2019JCAP...02..019N}.}
    \label{fig:spin1}
\end{figure}

The ML template presented in~\cite{2019JCAP...02..019N} should be accompanied
by a presentation of the distribution of the parameters in the MCMC chain. The
purpose is to check whether or not the values of the ML template, as well as
other templates with highest likelihoods, are consistent with the
distributions given by the whole converged MCMC chain. Such a test indicates,
that the ML template is very special in the MCMC chain. By simply reading the
37240 points of the chain used by~\cite{2019JCAP...02..019N}, we have
constructed the scatter plot of the spin1 amplitude versus log-likelihood
shown in Figure~\ref{fig:spin1}. One can see that most of these templates have
low spins. However, the ML template has extreme spin of $a_1=0.977$ that
almost hits the boundary of allowed priors that ranges from $0$ to
$0.99$~\cite{2018arXiv180710312B}, and is greater than 99.6$\%$ of the chain
templates, i.e., the parameters given by ML template deviate significantly
from those of the main chain. A similar situation is seen for the celestial
longitude (ra), which is also shown in Figure~\ref{fig:spin1}. Moreover, both
phenomena can be seen in the new chain as well.

All templates shown in Figure~\ref{fig:spin1} have relatively high
likelihoods, and the templates with highest likelihoods (including the ML
template) are expected to have a posterior distribution consistent with that
of the main chain. If the above high-spin anomaly were found for the single ML
template, then one may consider solving the problem by excluding this
template. However, this is not the case: For example, all 8 templates of
highest likelihoods have spins that are in the top 12.5$\%$ (see the colored
crosses in Figure~\ref{fig:spin1}), which has a probability of
$\approx6\times10^{-8}$. For the 1024 templates of highest likelihoods, 231 of
them are in the same region, which corresponds to a probability of
$\approx2\times10^{-19}$. Thus the inclusion of more templates with highest
likelihood quickly sharpens this problem rather than alleviating it. This
raises concern about the approach of~\cite{2019JCAP...02..019N}, not only for
using a single ML template, but also for the correctness of favoring higher
and higher likelihoods in their MCMC realization.

\subsection{Degrees of freedom, fluctuations in the SNR, and residual
correlations}\label{sub:cons}

In~\cite{2019JCAP...02..019N}, each point in the MCMC chain corresponds to one
template. The 15th parameter of the template (see Table~\ref{tab:alex params})
is determined by matching to the strain data with all other parameters fixed.
However, a proper matching should contain 6 degrees of freedom (DOF):
position, amplitude, and phase for each detector. As a quick consistency
check, we continue to use the code of~\cite{2019JCAP...02..019N} but add a
full matching of all 6 DOF using their pyCBC package. Our aim is to focus only
on the change from 1 to 6 DOF. The corresponding residual correlations for all
37240 templates in the 16.39s -- 16.43s window are plotted in the left panel
for 1 DOF and right for 6 DOF. We see that, a partial 1 DOF matching (as
adopted in~\cite{2019JCAP...02..019N}) leads to residual correlations with
large fluctuations even though all templates are within $\pm0.5\%$ relative
variation of the SNR. The range of fluctuations is significantly reduced by a
full 6 DOF matching but is still large. Therefore, the templates in the MCMC
chain tend to produce residual correlation with fluctuations that are large in
comparison with the small variations in SNR. Such large fluctuations may not
be physically meaningful.

\begin{figure}[!h]
\begin{center}
    \includegraphics[width=0.48\textwidth]{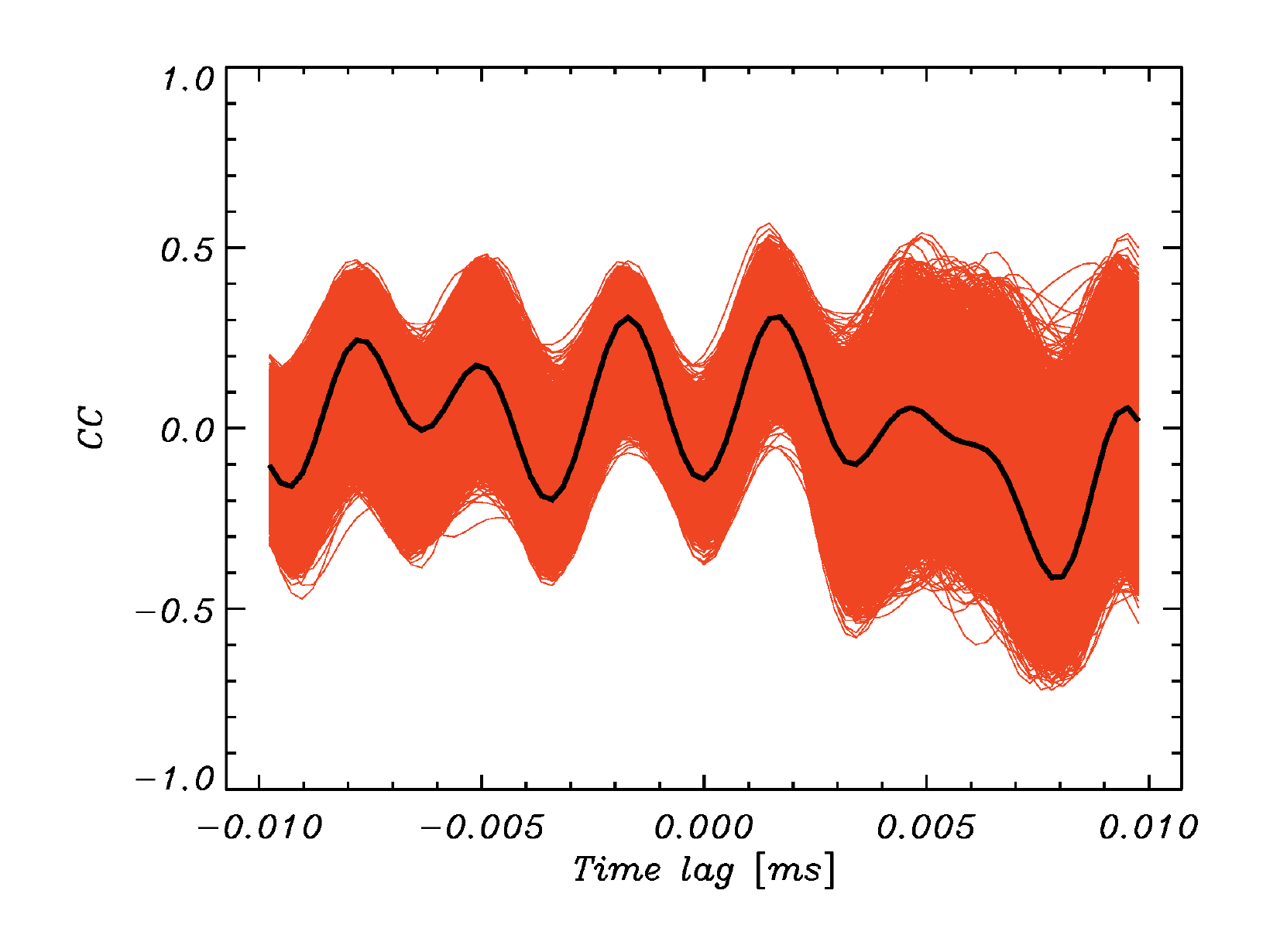}
    \includegraphics[width=0.48\textwidth]{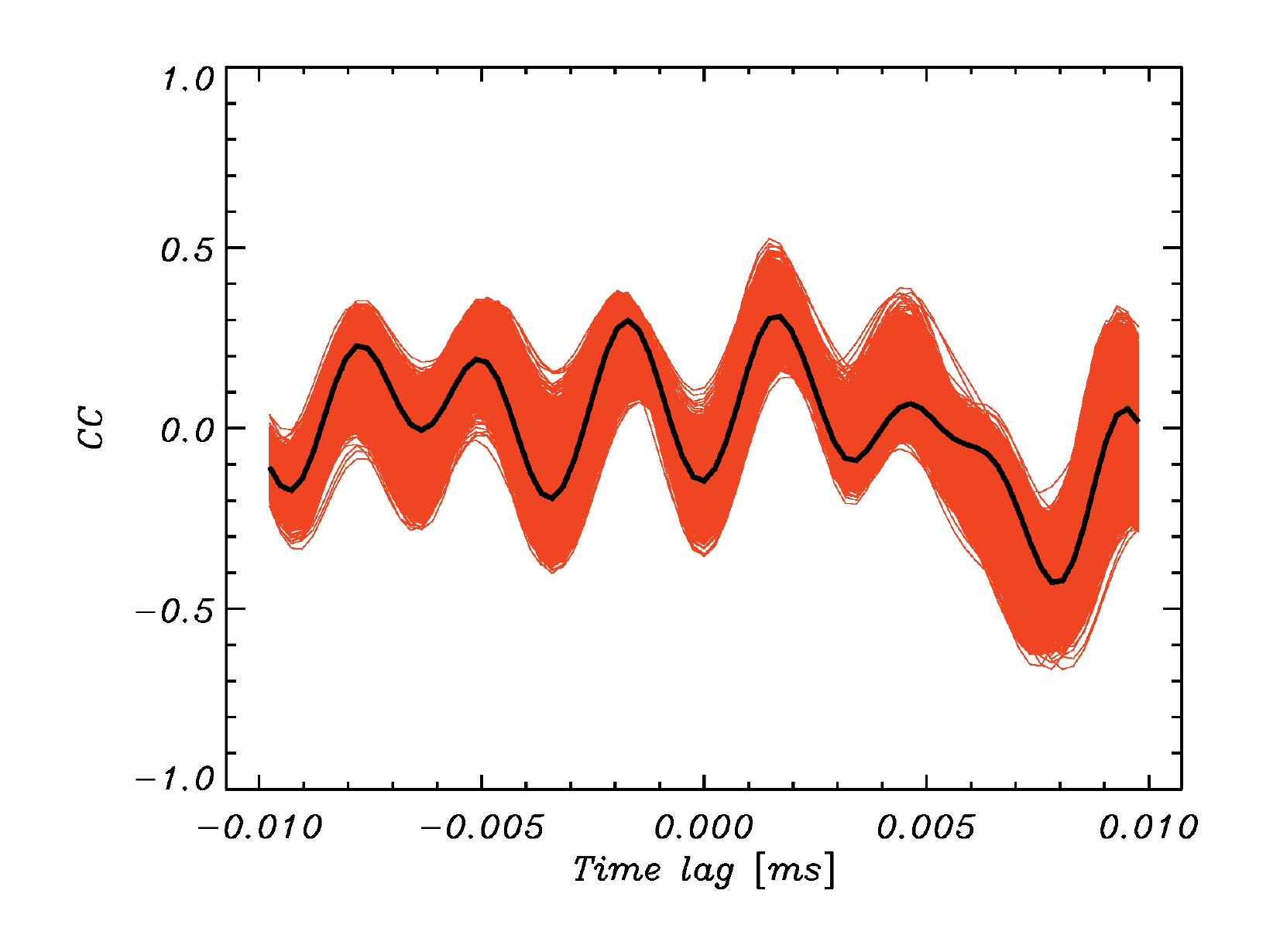}
\end{center}
    \caption{\emph{Left}: Residual correlations for region 16.39s--16.43s for
    all 37240 templates in the MCMC chain. The one for ML template in black.
    \emph{Right}: Same as left but each template is matched by 6 DOF instead
    of 1 DOF.}
    \label{fig:spin1b}
\end{figure}

\section{Hanford-Livingston noise residuals and their
cross-correlations}\label{sec:HL residual and cc}

In Table~\ref{tab:alex params}, 8 of the 14 parameters (except the coalescence
phase) describe intrinsic properties of the Black Hole binary. These include
the mass, spin amplitude, and polar/azimuthal angles for each black hole. The
remaining 6 parameters are related to the projection of GW onto detectors. A
straightforward matching for 2 detectors will also use 6 parameters including
the arrival time, amplitude and phase at each detector. These 6 parameters can
be translated into the GW projection parameters. Therefore, for 2 detectors,
either way (ML or straightforward matching) has the same number of free
parameters for the GW projection. For $n>2$ detectors, a straightforward
matching requires more parameters than there can be, and the projection
procedure must be constrained properly. Thus, a ML approach using 15
parameters can be more convenient for $n>2$ detectors even if each point in
the chain is unmatched. However, for two detectors like GW150914, there is no
such constraint. Each point in the MCMC chain is ``unmatched'' and should be
``rematched''.

Before proceeding to the residual correlations, it is useful to look at two
distinct methods for matching GW templates to the strain data. The most
important consideration for the rematching of templates is the
non-stationarity of the GW signal. For the template parameters given in
Table~\ref{tab:alex params}, the frequency of the GW emitted during the
inspiral exceeds the low frequency of the band-pass filter $f_{\rm min}\simeq
30$\,Hz for a mere $\approx 0.15$\,s before the chirp. After the chirp, both
the ML and NR templates fall rapidly down to the noise level. The result is
that the GW150914 signal is visible only for approximately $0.15-0.2$\,s. A
time interval of 0.2\,s corresponds to a frequency resolution 5\,Hz. Such poor
resolution is not sufficient to ensure a proper match of the template in the
frequency domain. In contrast, rematching techniques, based on
Eq.~(\ref{corr}), that maximize the cross-correlation between the strain data
and the template for a short 0.2\,s record are unaffected by the this poor
frequency space resolution.  In the discussions below we will consider both
templates that are rematched using the Fourier methods
of~\cite{2019JCAP...02..019N} and templates that we rematch directly in the
time domain as well as the corresponding residuals.

The use of a PSD obtained from 1040\,s of data~\cite{2019JCAP...02..019N} in
order to determine the parameters of a template that is only visible for
0.2\,s obviously requires the assumption that the noise is stationary.  This
assumption is known to be violated, and the noise is tainted by non-Gaussian
features and non-stationary behavior. In the presence of such deviations from
ideal noise, a more conservative approach is to restrict the analysis to short
segments in the time domain within which the effects of non-stationarity and
non-Gaussianity are potentially less damaging.  After bandpassing and
notching, the template is largely localized in 200\,ms region.  We use data
from this region alone to perform rematching of the template to the data.

\begin{figure}
    \centerline{
    \includegraphics[width=0.48\textwidth]{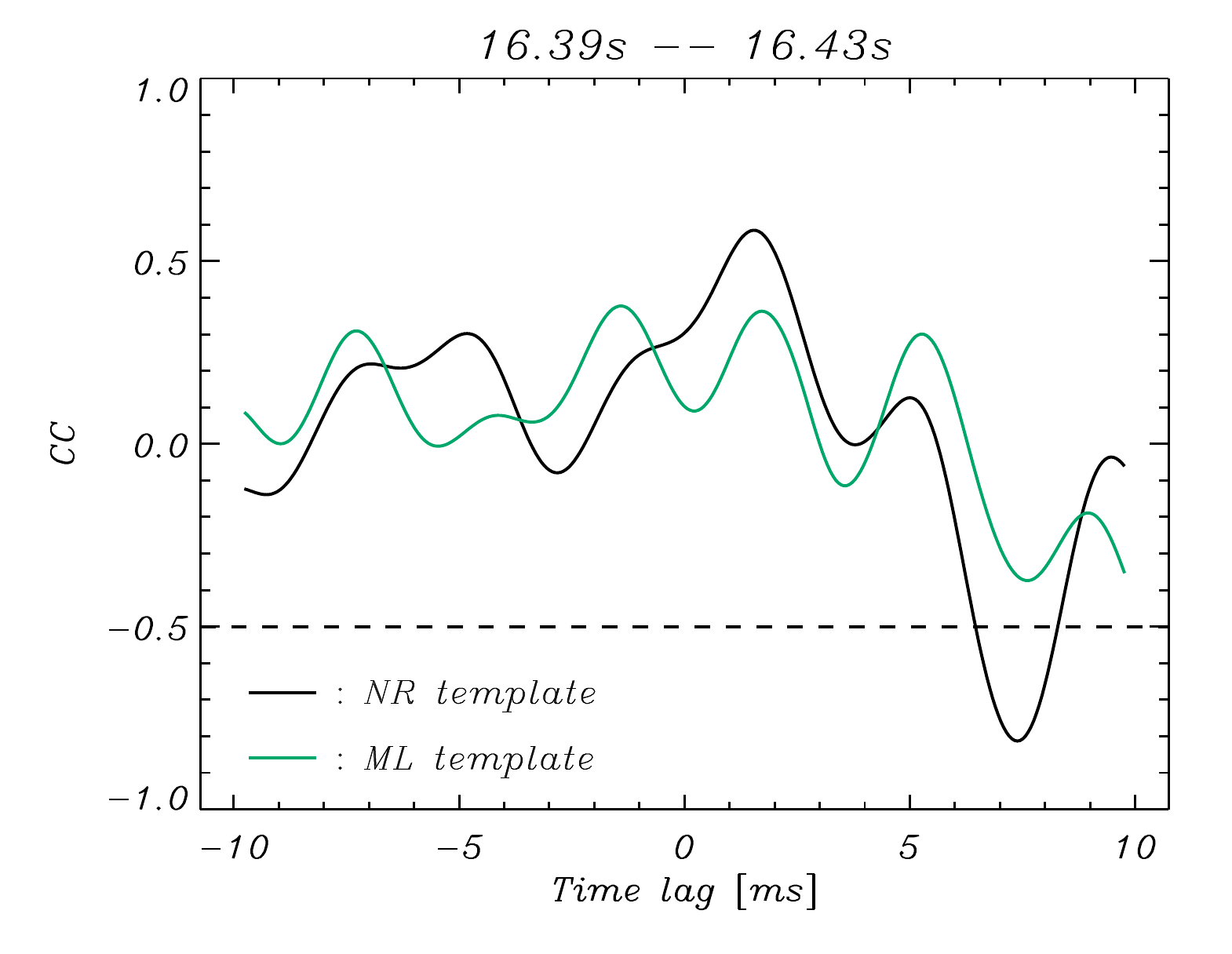}
    \includegraphics[width=0.48\textwidth]{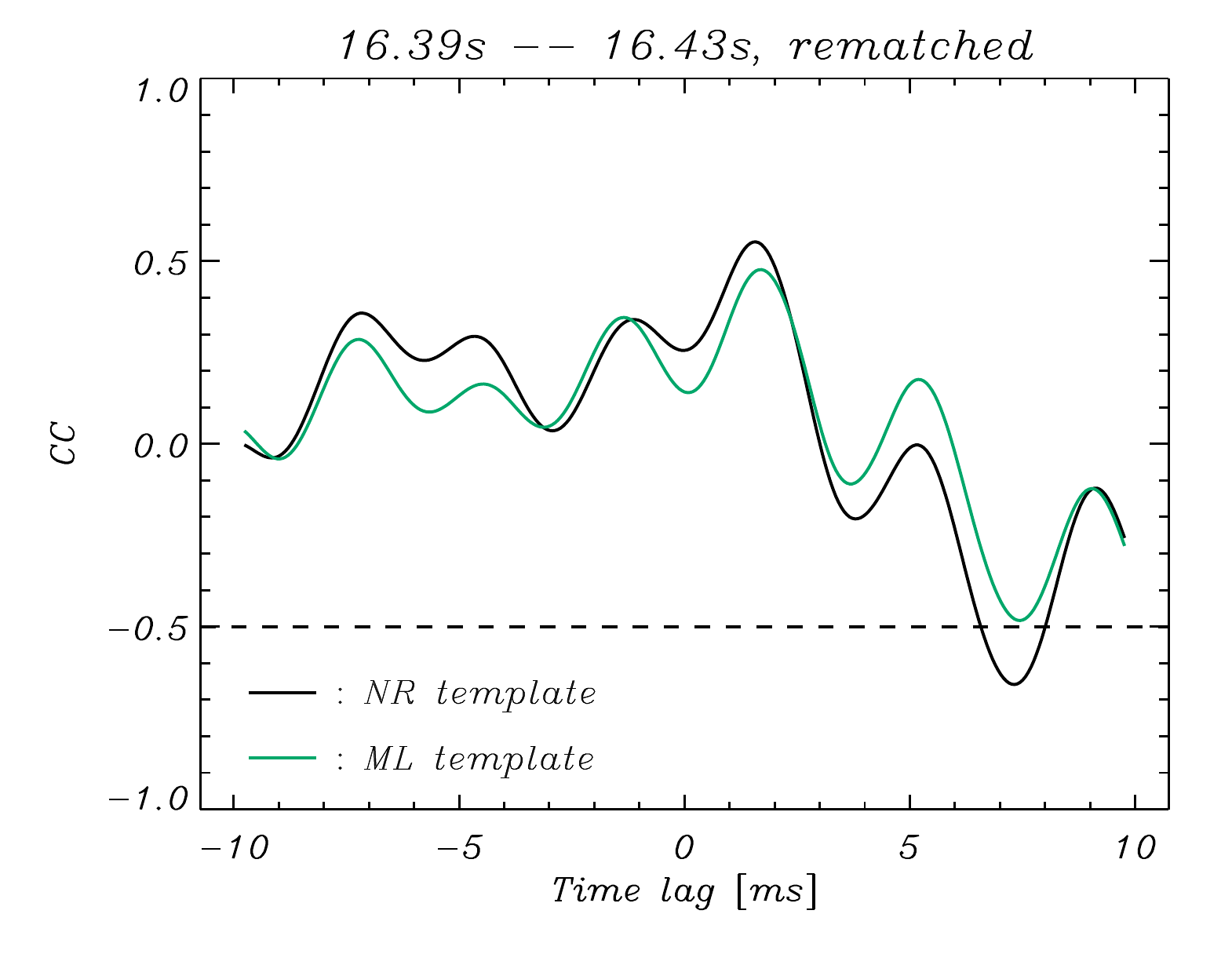}
    }
    \caption{Left panel: Comparison of the ML template residual
    cross-correlation with that for the NR template.  The green curve is
    identical to Figure 2 in~\cite{2019JCAP...02..019N}.  Right panel: The
    same residual cross-correlations after rematching.  The black and green
    curves are quite similar; their cross correlation between the black and
    green curves is 0.95. }
    \label{fc1}
\end{figure}

In Figure~\ref{fc1} we summarize the results presented
in~\cite{2019JCAP...02..019N} for the cross correlation of the Hanford and
Livingston residuals using their publicly available program. Results are also
shown for the NR template in order to provide a more complete picture of cross
correlations in the 40\,ms time domain surrounding the chirp. This figure also
shows the results after rematching both templates to the data as follows.
\begin{itemize}
\item The rematching procedure is intended to maximize separately the
correlations defined in eq.~(\ref{corr}) between the Hanford and Livingston
strain data and the template.
\item The rematching is performed in the 16.25-16.45\,s time window with a
sampling rate of 16\,kHz for all templates considered\footnote{Note that the
pixel size for the ML template is 0.25\,ms. This is why we perform the
rematching using a 16\,kHz sample rate.}.
\item The amplitudes and the phase, $\phi_0$, are not fixed; they are to be
determined by optimization.
\item The results of Hanford rematching are a pixel shift of 0 pixels, a phase
correction of $\Delta\phi_0 \approx -0.05$, and an amplitude correction factor
of 1.0241.
\item The resulting of Livingston rematching are a pixel shift of -2 pixels
(-0.125\,ms), a phase correction of $\Delta\phi_0 \approx -0.079$, and an
amplitude correction factor of 1.0236.
\end{itemize}
After rematching, we check the cross-correlation coefficient between the
bandpassed and notched strain and the template in the  16.25--16.45s window
and find an increase from 0.893 to 0.894 for Hanford and an increase from
0.842 to 0.844 for Livingston, which is consistent with our expectation. For
Gaussian random noise, which provides the basis for the ML approach, a higher
cross-correlation coefficient necessarily means a higher SNR.

\begin{figure}
\centerline{
\includegraphics[width=0.8\textwidth]{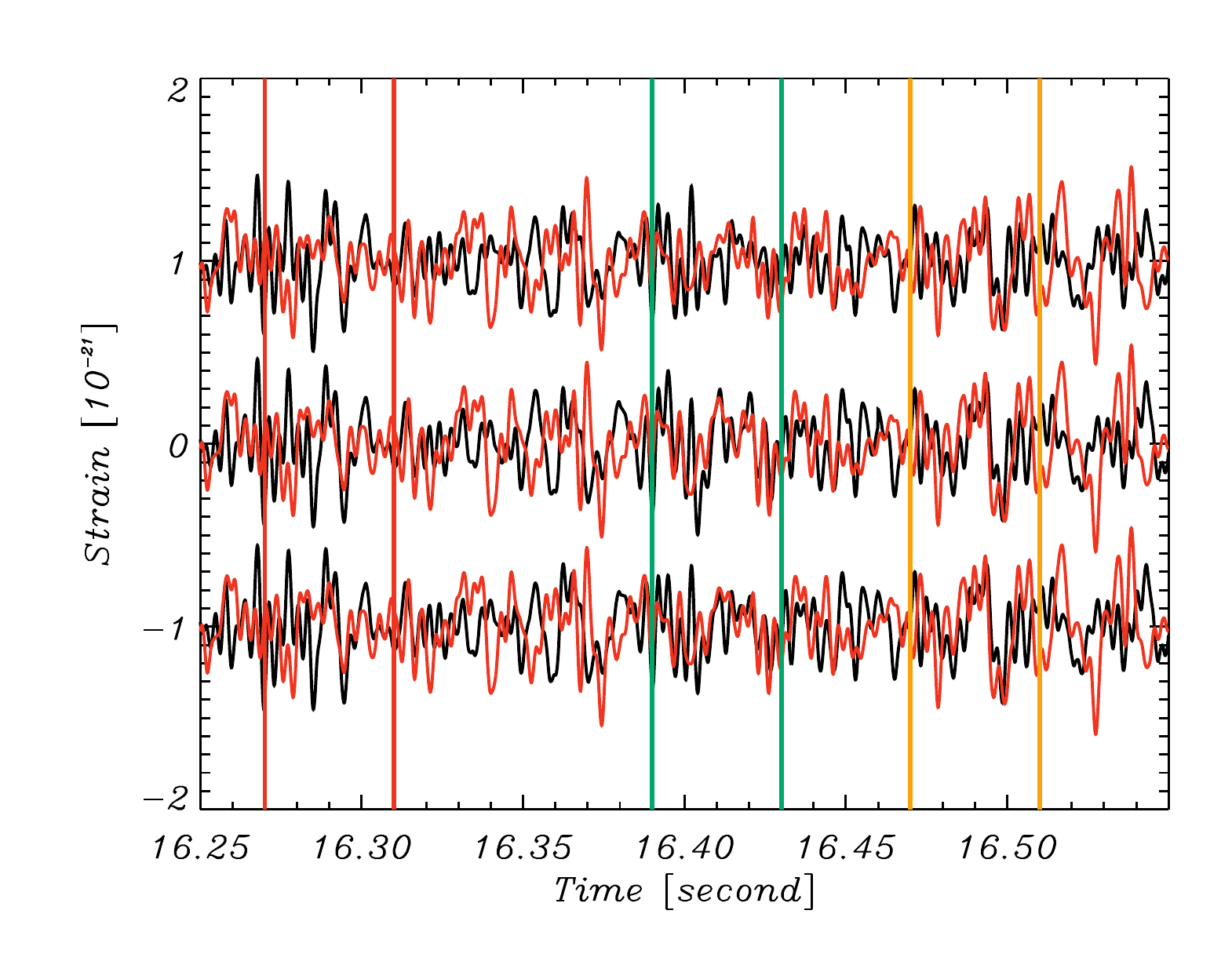}
}
\caption{The Hanford (black) and Livingston (red) residuals for the ML
  template (top), NR template (middle), and the rematched NR template
  (bottom). The 16.39--16.43 region (green), the precursor region (red) and
  the echo domain (yellow) are also shown. (See Appendix G and Figure~24
  of~\cite{2017JCAP...08..013C} for details.) Note that Livingston residuals
  have been shifted by 7\,ms and inverted. All residuals have been bandpassed
  into the frequency range $35\le f\le 350$ Hz. }
\label{figc2}
\end{figure}

\begin{figure}
\centerline{
\includegraphics[width=0.33\textwidth]{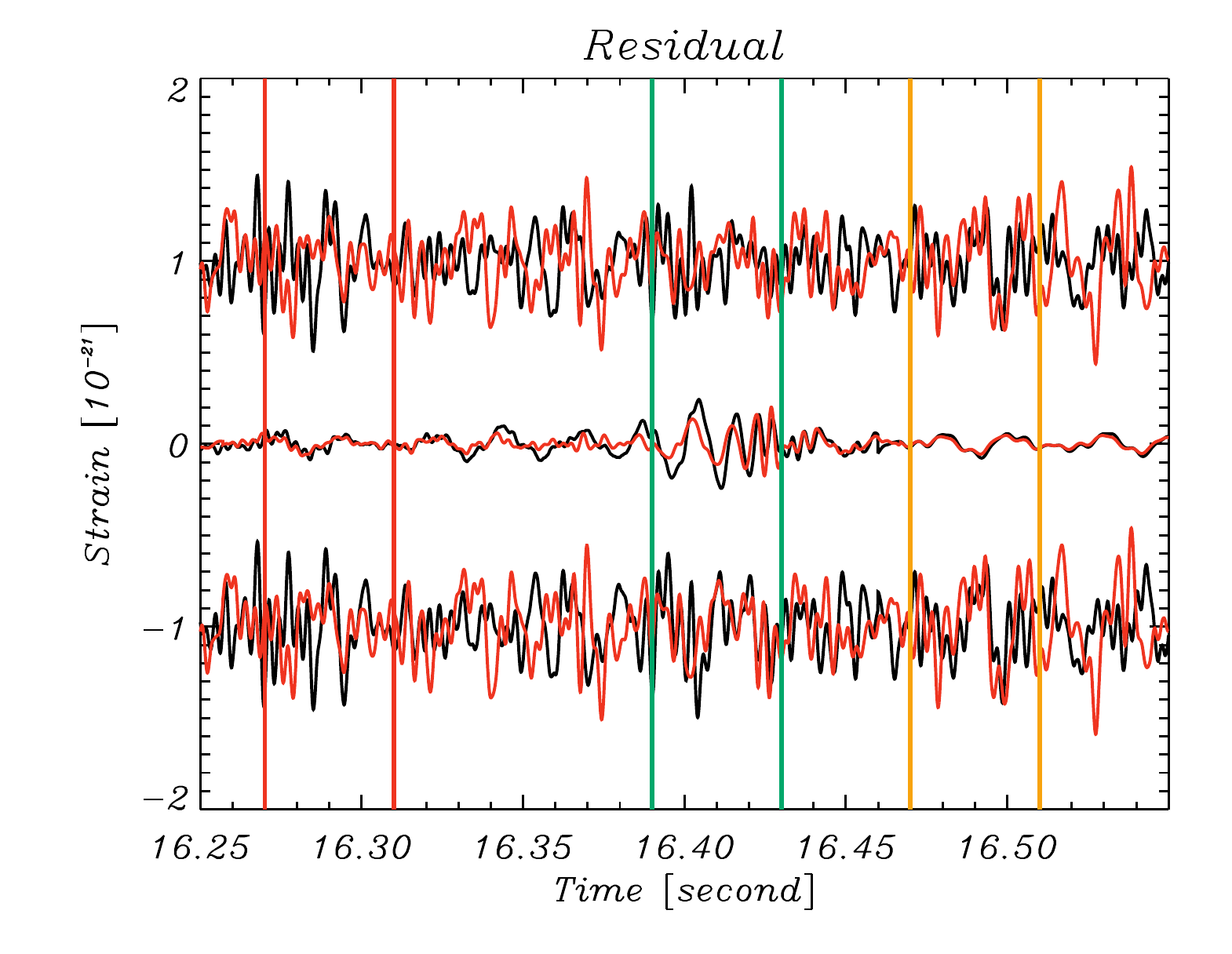}
\includegraphics[width=0.33\textwidth]{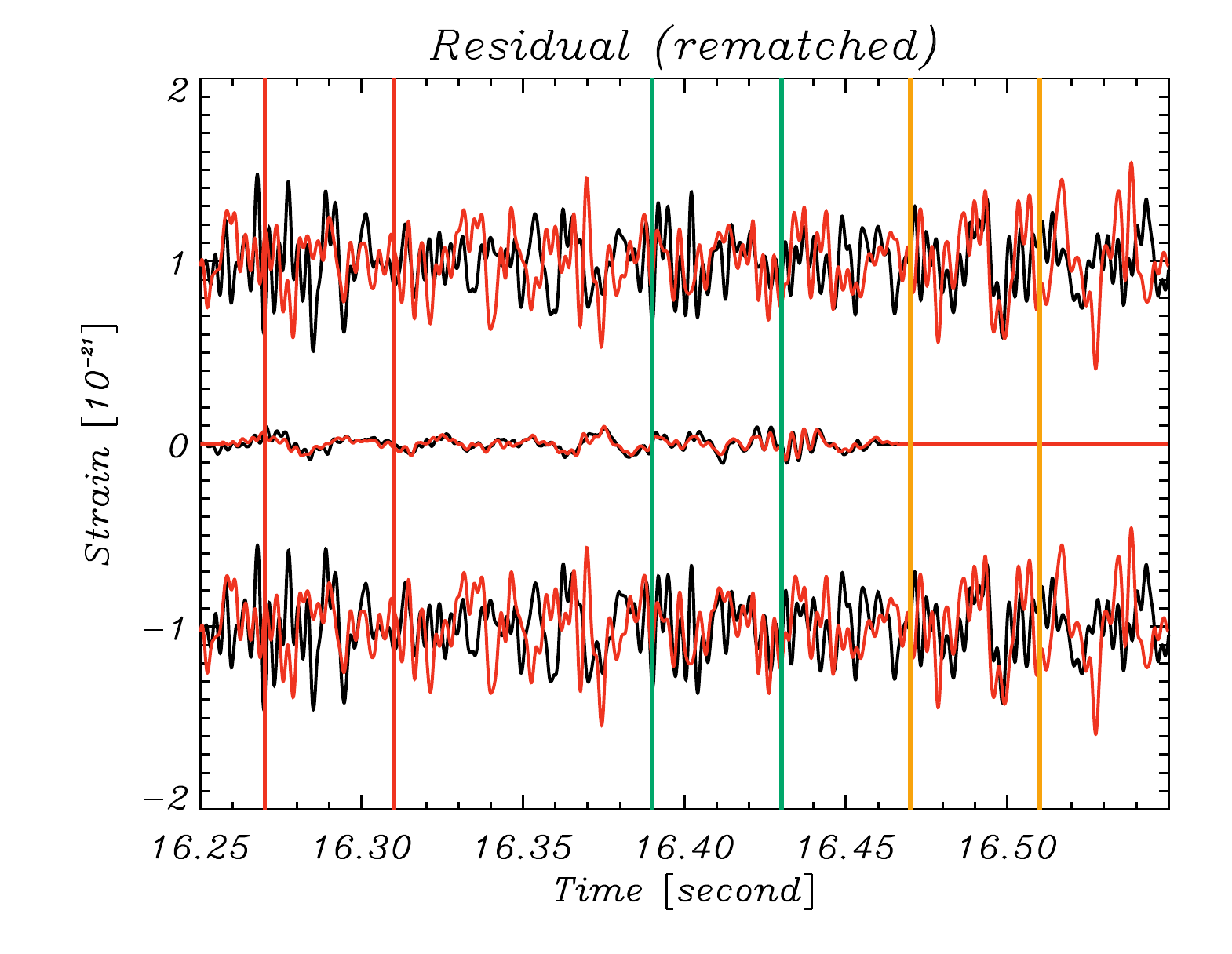}
\includegraphics[width=0.33\textwidth]{CC2.pdf}}
\caption{Left panel:  The Hanford (black) and Livingston (red) residuals
  without rematching.  The upper plot is for ML and the lower for NR. The
  middle curve is the difference between them.  Middle panel: The same as the
  left panel but for rematched templates. Livingston residuals have again been
  shifted by 7\,ms and inverted. All residuals have been bandpassed into the
  frequency range $35\le f\le 350$ Hz. Right panel: the H/L cross-correlation
  coefficients in the 40\,ms chirp domain after rematching. The same as the
  right panel of Figure~\ref{fc1}. Black for ML and green for NR. The
  cross-correlation between the black and green curves is 0.95.}
\label{figc3}
\end{figure}

Figures~\ref{figc2}--\ref{figc3} show the H/L residuals resulting after
templates have been subtracted from the strain data. Results are shown for the
ML, NR and rematched NR templates.  All figures confirm our expectation that
the correlations in the precursor and echo time domains are insensitive to the
choice of template. As shown in~\cite{2017JCAP...08..013C}, these regions are
characterized by cross-correlation coefficients on the order of $0.6$--$0.7$.
In the chirp domain the structure of the residuals for the rematched ML and
rematched NR templates are more similar in spite of the fact that they are
significantly different for the non-rematched case.

\section{Evaluation of the significance of residual
correlations}\label{sec:resi-CC significance}

In our view, the significance estimator used by~\cite{2019JCAP...02..019N} is
both biased and counterintuitive. In Section~\ref{sub:more robust}, we explain
why this is so and use a simple but more robust estimator to calculate the
significance.  In Section~\ref{sub:with old estimator} we then use the
estimator of~\cite{2019JCAP...02..019N} with necessary corrections to obtain
results that can be compared with the results of~\cite{2019JCAP...02..019N}.
Moreover, by comparing the significances estimated in Section~\ref{sub:more
robust} and Section~\ref{sub:with old estimator} (with corrections), we see
that they in good agreement and that information regarding the peak position
is of dominant importance.  This will be further illustrated in
Appendix~\ref{app:toy model}.

\subsection{More robust estimation of the significance of residual
correlation}\label{sub:more robust}

The most important feature of the residual correlation is that it appears with
a time lag identical to that of the GW event itself. Since the physically
allowed time lag range is $\pm 10$ ms, a robust estimation of the probability
of the ``time lag coincidence'' is simply\footnote{Strictly speaking, the
$p$-value given by eq.~(\ref{equ:simple p}) should be combined with the
amplitude of the lowest point to give a complete estimation. However, we have
confirmed that this introduces only minor corrections of relatively 10--20\%,
e.g., a change 4\% to 3.6\%. Thus, for simplicity, we ignore this correction
here.}
\begin{equation}\label{equ:simple p}
    p = 2\Delta t/20,
\end{equation}
where $\Delta t$ (in ms) is the distance between the positions of the lowest
points of the residual and template cross correlations. Eq.~(\ref{equ:simple
p}) also has the considerable advantage that it is correct even for
non-stationary and non-Gaussian noise provided only that noise in the Hanford
and Livingston noises is independent. To illustrate the stability of this
estimator, we observe that in Figure~\ref{fig:spin1b}, 99.8$\%$ of the
rematched residual correlations (right panel) have their lowest points
clustered between 7-8 ms.  Even without rematching (left panel), 75$\%$ of
them are still clustered in this range.

For the case of the ML template without rematching, $\Delta t$ is less than
0.4\,ms. Thus, from eq.~(\ref{equ:simple p}), the probability of ``time lag
coincidence'' is $4\%$. When rematching is performed, $\Delta t$ is reduced to
0.06\,ms, and the probability is less than $1\%$. However,
in~\cite{2019JCAP...02..019N} the probability is given at about $40\%$, which
is completely counterintuitive. This strongly suggests that the estimator used
in~\cite{2019JCAP...02..019N} is significantly biased.

In Figure 5 of~\cite{2019JCAP...02..019N}, the authors performed an arbitrary
modification of the chirp window with the aim of demonstrating that the
residual correlation is insignificant. Such an arbitrary choice is not
sufficient to support their claim.  However, it is still useful to consider
modifications of the window chosen.  A more reasonable test of this kind can
be carried out as follows. We allow the window to start at 16.38, 16.39 or
16.40 seconds, and to end at 16.43, 16.44 or 16.45 seconds.  The combinations
give 9 different windows at various positions and with various lengths between
30 and 70\,ms. Residual correlations have been calculated for all 9 cases
(yellow) as well as their average (black), as shown in Figure~\ref{fig:9
cases}. Apparently this 9-window average estimation is much more reliable than
the arbitrary choice in~\cite{2019JCAP...02..019N}. Before rematching, $\Delta
t$ is found to be about 0.2\,ms, corresponds to $p=2\%$, whereas after
rematching, $\Delta t$ is less than 0.06\,ms, corresponding to $p<1\%$. Later
in Section~\ref{sub:9 window for old estimator}, the same 9-window case is
reevaluated using the corrected estimator of~\cite{2019JCAP...02..019N}, and
the same 1\% probability is found.
\begin{figure*}[!htb]
  \centering
  \includegraphics[width=0.48\textwidth]{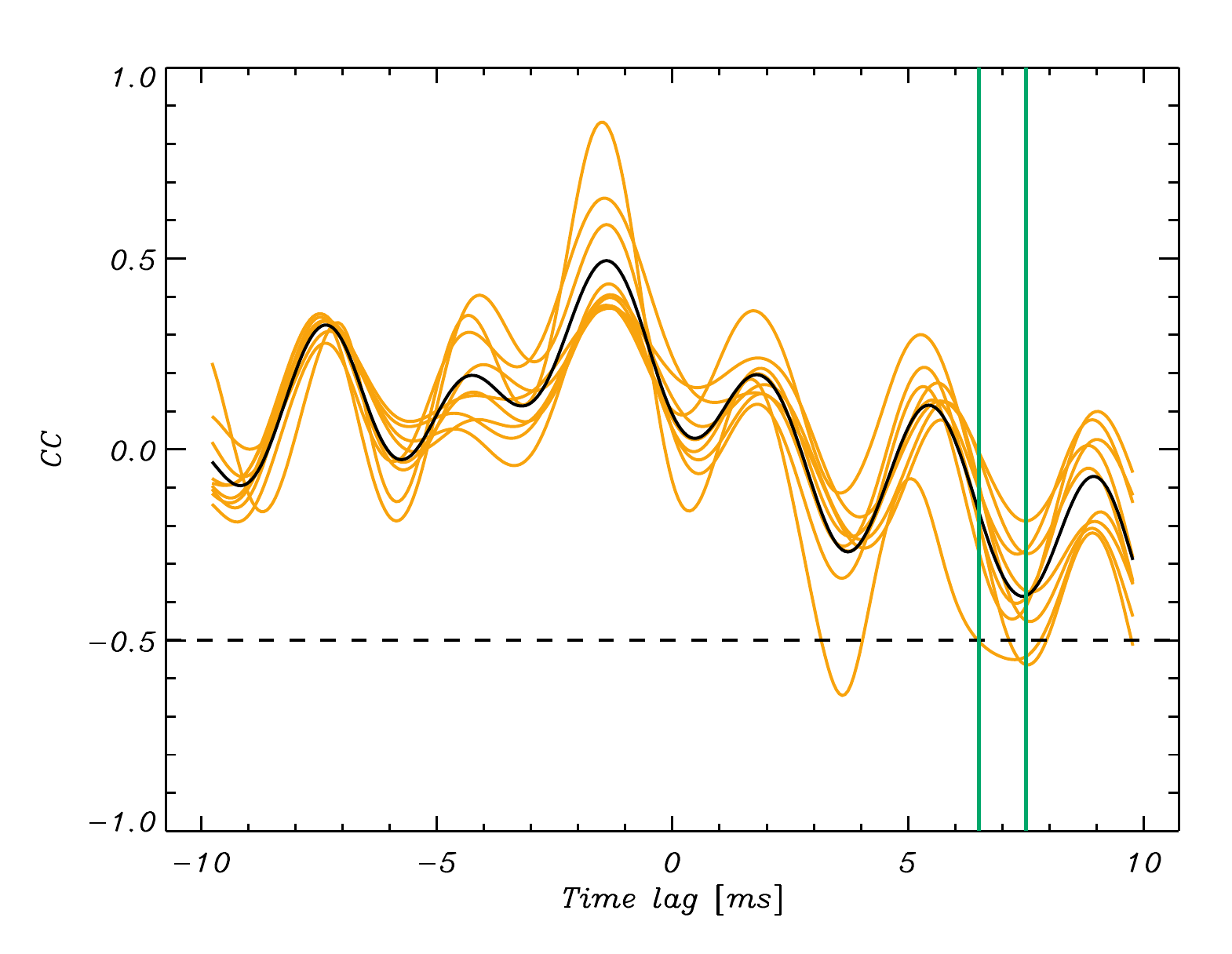}
  \includegraphics[width=0.48\textwidth]{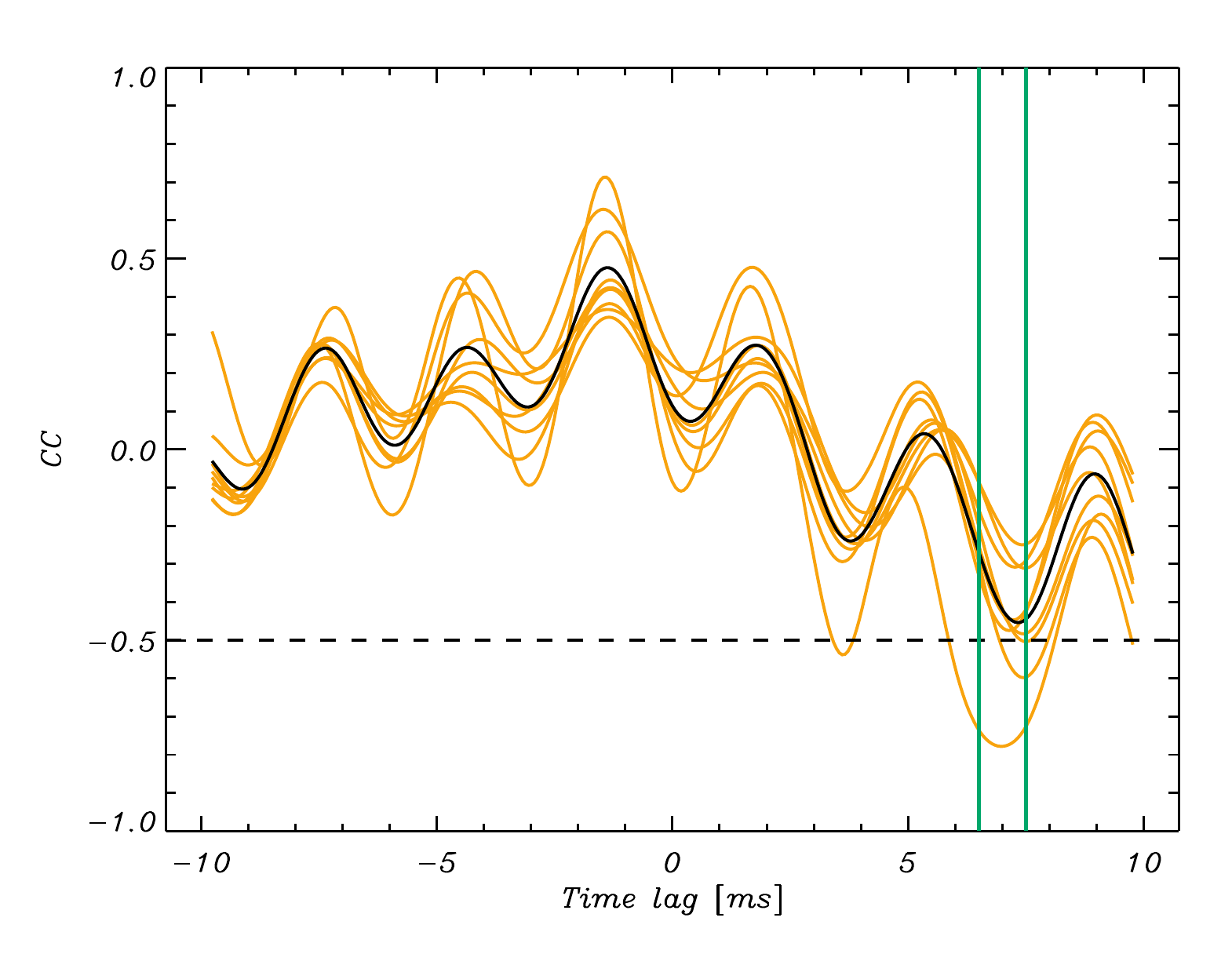}
  \caption{Residual correlations for 9 windows with the windows
  starting from 16.38, 16.39 and 16.40s, and end at 16.43, 16.44, 16.45s,
  respectively. The average of 9 windows is plotted in black. Left: no
  rematching, right: with rematching.}
  \label{fig:9 cases}
\end{figure*}

\subsection{Calculations with the old estimator}\label{sub:with old estimator}

In the event that one elects to use the significance estimator
of~\cite{2019JCAP...02..019N}, at least two adjustments are required. Without
such corrections, the results will be strongly biased. These include:
\begin{enumerate}
    \item To obtain a reliable p-value for the observed residual
    cross-correlation, a correction factor of roughly 0.5 should be applied
    because we are interested in the significance of the physically meaningful
    cross correlation including its sign and not the significance of the
    absolute value of the cross correlation as in~\cite{2019JCAP...02..019N}.
    \item To obtain a reliable p-value for the observed residual
    cross-correlation, at least two points should be considered: a) The local
    minimum in the 6.5--7.5\,ms window as given by simulation is lower than
    that obtained with real data, and b) The global minimum in the entire
    $\pm10$\,ms range lies in the same 6.5--7.5\,ms window.  Unfortunately,
    in~\cite{2019JCAP...02..019N} only point a) is considered and b) is
    completely ignored. This provides another indication that these results
    are biased.
\end{enumerate}
After considering both corrections, the resulting probability, even with the
estimator of~\cite{2019JCAP...02..019N}, will decrease by roughly one order of
magnitude and become consistent with the results obtained in
Section~\ref{sub:more robust}.

In fact, even the choice to use the 6.5--7.5\,ms is based on the a priori
assumption that one already knows that the true time lag is 7\,ms. A more
natural choice might be to consider $\Delta t$ as used in eq.~(\ref{equ:simple
p}).

\subsubsection{Example of varying the window position}
\label{sub:The chirp and precursor regions}

\begin{figure*}[!htb]
  \centering
  \includegraphics[width=0.48\textwidth]{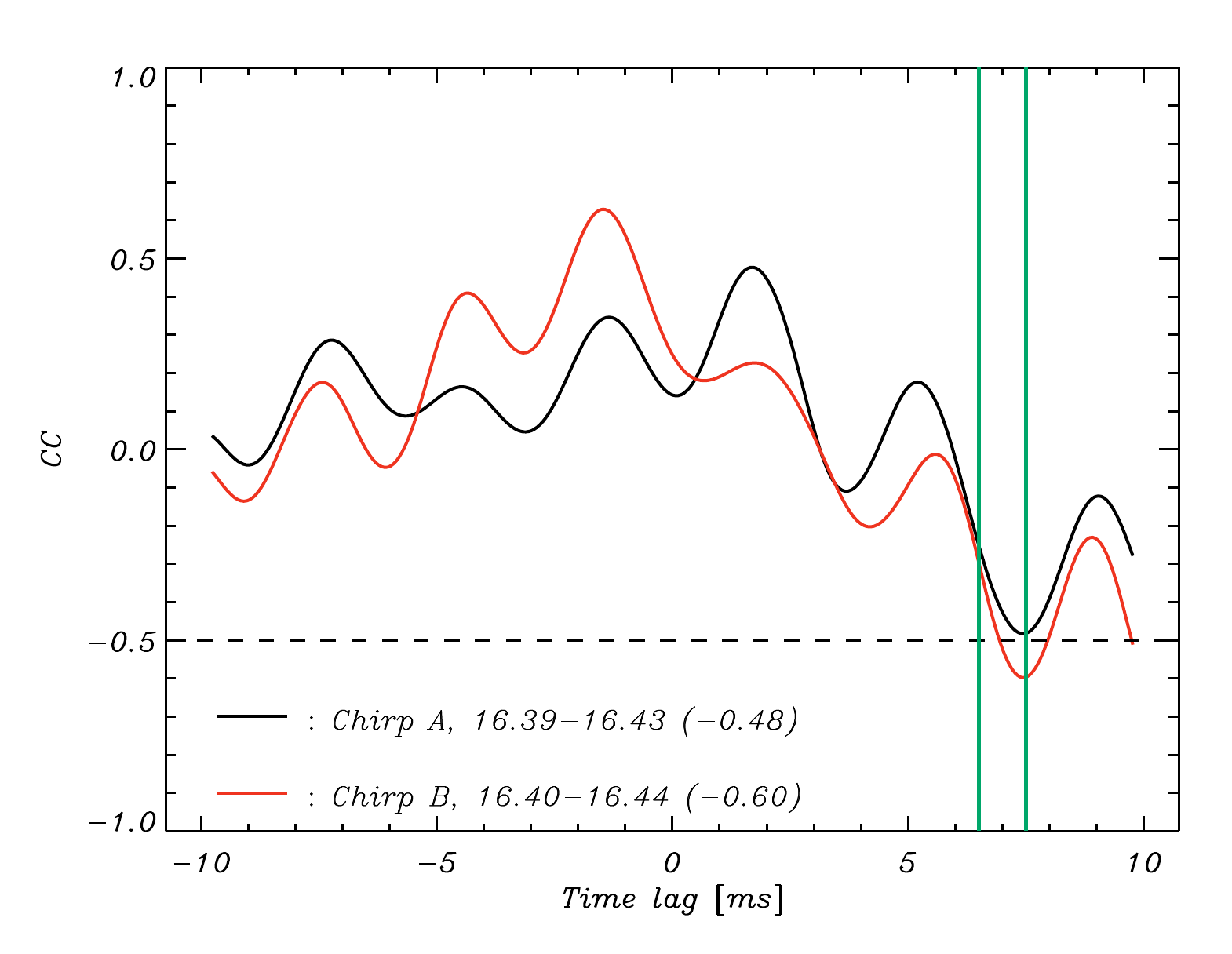}
  \includegraphics[width=0.48\textwidth]{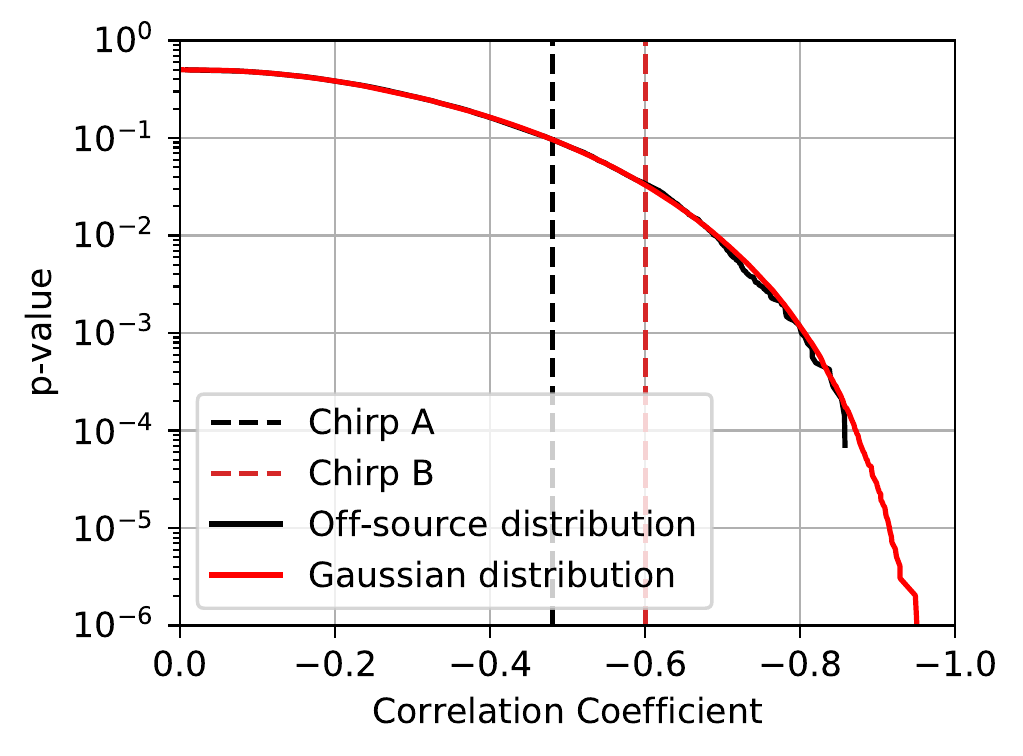}
  \caption{Left panel: The H/L residual correlation using the rematched
  templates and   residuals for the chirp A position  (16.39s--16.43s) and the
  chirp B position (16.40s--16.44s).  Right panel: The p-value as a function
  of the residual correlation reproduced from~\cite{2019JCAP...02..019N} with
  the inclusion of the proper factor of $1/2$ as discussed in the text. The
  vertical lines indicate the cross-correlations found for chirp A and chirp
  B.}
  \label{fig:pos ABC}
\end{figure*}

In our previous work~\cite{2017JCAP...08..013C}, the window used for the
calculation of the residual correlation was 16.39--16.43s. However, we also
noticed that this position is somewhat more sensitive to uncertainties in the
choice of template than a slightly shifted window running from 16.40--16.44s.
For convenience, we will refer to the original window as ``chirp A'' and the
second as ``chirp B''.  We have used the rematched ML template and residuals
to calculate the residual correlation for these windows, and the resulting
residual correlations are shown in Figure~\ref{fig:pos ABC} where the maximum
anticorrelations are seen to be -0.48 and -0.60 respectively.

In~\cite{2019JCAP...02..019N}, the authors presented a significance estimation
of the residual correlation by simulation. However, as mentioned above, they
consider only the amplitude but ignore the sign.  When the sign of the
correlation is correctly taken into consideration, the p-value given in their
Figure 3 should be multiplied by a factor of 0.5. We reproduce their Figure 3
as the right panel of Figure~\ref{fig:pos ABC} including this factor and
indicate the residual correlations found at the precursor, chirp A and chirp B
positions. We see that the residual correlations for the chirp B window has a
significance of $p_B\approx4\%$. If the position of the lowest point is taken
into account, the probability will further decrease.

\subsubsection{Significance estimation with 9 window average}
\label{sub:9 window for old estimator}

As mentioned above and in Section~\ref{sub:more robust}, one should allow the
window position and size to change in a way that is more reasonable than the
arbitrary choice in Figure 5 of~\cite{2019JCAP...02..019N}. In
Section~\ref{sub:more robust}, we used the average of 9 windows to give a more
robust estimation. Here we use the same 9-windows and the corrected estimator
as described at the beginning of Section~\ref{sub:with old estimator} to
re-calculate the significance. The resulting $p$-value is 1\%, which is
consistent with the result for 9-windows using eq.~(\ref{equ:simple p}).
Therefore, we conclude that, with the use of the ML template, the probability
of getting the observed residual correlation in the chirp domain by chance is
less than $1\%$.

\subsubsection{Investigation of the RMS peaks}\label{sub:precursor and joint}

In our previous work~\cite{2017JCAP...08..013C}, the positions of the
precursor and echo were detected with a running window calculation of the H/L
residual correlation. These features of the noise can also be verified by the
running window RMS test, defined in eq.~(\ref{equ:joint rms})
of~\cite{2017JCAP...08..013C}. Note that this RMS test is independent of the
test performed in~\cite{2017JCAP...08..013C}:  The RMS focuses on amplitudes;
the test performed  in~\cite{2017JCAP...08..013C} depends only on morphology.

\begin{figure*}[!htb]
  \centering
  \includegraphics[width=0.96\textwidth]{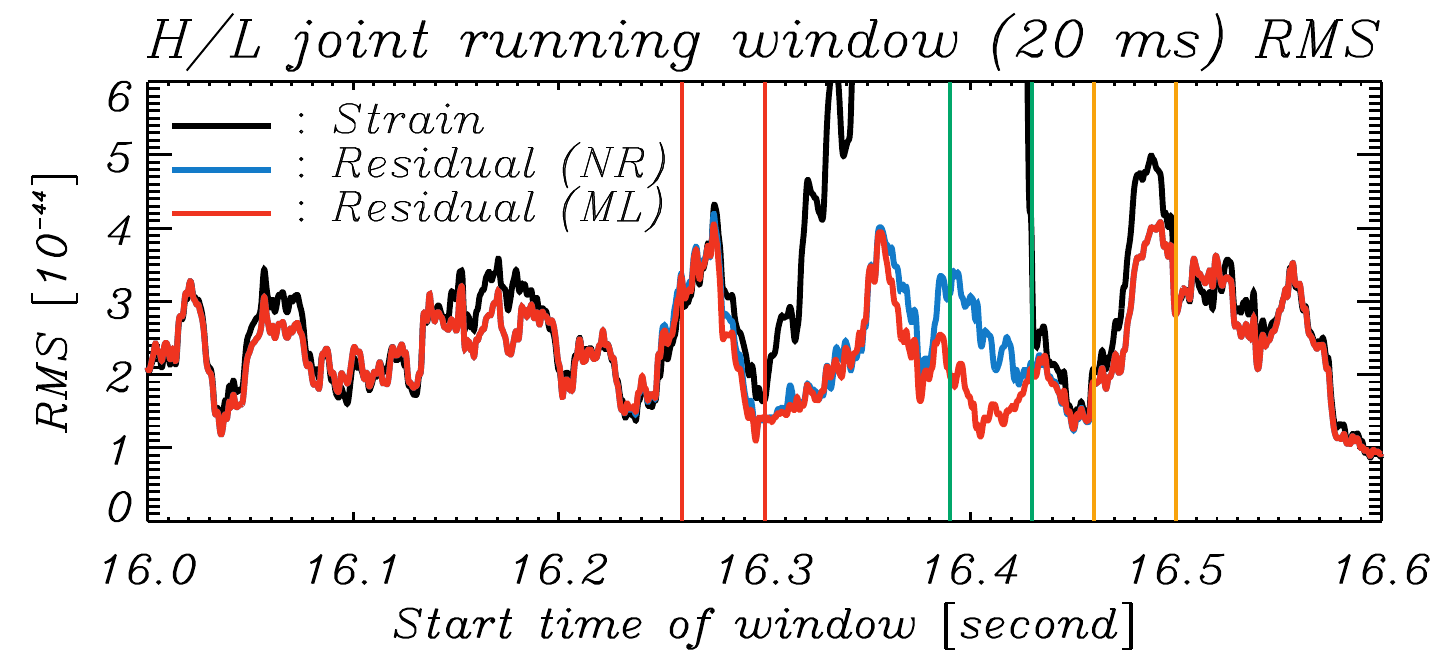}
  \caption{The Hanford and Livingston joint RMS $\delta$ calculated for 20\,ms
  running windows using eq.~(\ref{equ:joint rms}).  The black curve is for the
  H/L strain (template not removed), the red curve is for the residuals
  calculated with ML template, and the blue curve is for the residuals
  calculated with the NR template (neither rematched). The precursor and echo
  windows are indicated by vertical lines. Since the $x$-axis denotes the left
  boundary of the window, we shift the marks by a half-window to the left to
  represent the effective region more accurately.}
  \label{fig:run win rms}
\end{figure*}

The RMS test is performed in the following way. We take a 20\,ms running
window\footnote{For a 40\,ms window, 10\,ms are excluded from each side in
order to avoid edge effects when calculating the residual correlation.  The
remainder will thus be of 20\,ms.} along the Hanford data and the Livingston
data (shifted by 7 ms and inverted).  For each window, we calculate the joint
RMS as
\begin{equation}\label{equ:joint rms}
\delta(t) = \frac{1}{n}\sqrt{ \left(\sum_{i=1}^n H_i(t)^2\right) \cdot \left(
\sum_{i=1}^n L_i(t)^2 \right) },
\end{equation}
where $t$ is the start time of the window, $i$ is the index of the data points
$H_i(t)$ and $L_i(t)$, and $n$ is the total number of points in the window.
Figure~\ref{fig:run win rms} shows a plot of $\delta$ as a function of the
time, $t$, of the left boundary of the window.  From this figure we see that
the precursor and echo windows correspond not only to regions of morphological
similarity~\cite{2017JCAP...08..013C} but also (and independently) to local
regions of joint signal strength.  It should also be noticed that, even though
the GW event is effectively finished after 16.45 seconds, the bandpassed and
notched template continues to show some signal as can be seen from the
difference between the black and red lines in Figure~\ref{fig:run win rms}.
(This effect is not visible for the NR template because the length of this
template is only 260\,ms.) The difference between the red and blue lines in
the chirp domain is also noteworthy because it shows that, unlike the NR
template, the ML template ``absorbs'' much more power and introduces
significant inhomogeneities (i.e., very low amplitudes) in $\delta(t)$ in the
chirp domain. This effect can also be seen in Figure~\ref{f4} in which
comparable results are obtained when the difference between the two GW
templates is rescaled by a factor of 5.

\section{Uncertainties in the maximum likelihood method}\label{sec:uncertain} 

It is obvious that no fit to data can be considered to be complete until a
reliable analysis of the associated uncertainties has been performed.  The
importance of such an analysis is even  greater when the resulting parameters
are presumed to have physical significance.  Consider, for example,
Figure~\ref{fig:spin1} that shows the amplitude of the spin of the larger-mass
black hole (upper panel) and the right ascension angle, ra.  The former is an
intrinsic property of a BH system; the latter determines the sky location of
the presumed BBH system.  The scatter plots show that each of these parameters
can assume essentially any of its possible values.  The single point of
maximum likelihood selects only one point.  The questions are thus: What are
the best values for these parameters, what are their uncertainties, and are
these uncertainties correlated?  Since~\cite{2019JCAP...02..019N} does not
address these issues, we will try to suggest some answers here.

It is important to understand the relation between the strain data ($s$), the
GW signal ($h$), and the noise ($n = s - h$).  The SNR, $\rho$, used
in~\cite{2019JCAP...02..019N} is given as
\begin{equation}
    \rho = \frac{\langle s|h \rangle}{\sqrt{\langle h|h \rangle}}
\end{equation}
with 
\begin{equation}
    \langle s|h \rangle = 4 \Re \int \frac{s^*(f) h(f)}{S_n(f)} df.
\end{equation}
Here, $S_n (f)$ is the power spectral density of the noise.  The
log-likelihood for Bayesian inference used by LIGO is
\begin{equation}
    \log(L) = -\frac{1}{2} \langle s-h|s-h \rangle, 
\end{equation}
which can be converted to $\rho^2/2$ with a proper offset. (The horizontal
axis in Figure~\ref{fig:spin1} is $\rho^2/2$.) It was
shown~\cite{PhysRevD.85.122006} that the variance of the log-likelihood
associated with different noise realizations is 1.\footnote{This derivation
makes the assumption of colored Gaussian noise, but the result has been
validated empirically as a good approximation for real detector noise. More
seriously, the derivation also assumes that $\langle n | h \rangle = 0$, which
is strictly valid only for the maximum likelihood template, and constitutes a
``higher-order'' correction otherwise.} However, in reality, the
log-likelihood should be treated with care. For example, $75\%$ of the points
in the MCMC chain have higher SNR than 24, the officially reported SNR of
GW150914~\cite{PhysRevLett.116.061102}, thus there is no reason to reject any
them. Meanwhile, in the new MCMC chain (see Table~\ref{tab:alex params}
and~\cite{gw150914:mcmc:ml:online}), the mean log-likelihood is lower than the
mean of the old chain by 2. Such a difference is about 200 time bigger than
the expected fluctuation by chance, assuming Gaussian stationary noise.
Finally, as pointed out in Figure~\ref{fig:spin1}, a few points look
suspicious, and they are unfortunately the ones with highest likelihoods.

In summary, except for some suspicious points, the majority of the MCMC chain
should be either all rejected, or all accepted. There is no reason to regard
the single ML template as best or most reliable, especially when it and its
kind show significant differences from the remainder of the chain. Rather,
each of the 37,240 templates represented in Figure~\ref{fig:spin1} is an
equally valid best-template candidate.  Thus, the results of
Figure~\ref{fig:spin1} and Figure~\ref{fig:spin1b} provide an illustration of
the uncertainties that should be associated with parameter determination and
residual correlations (respectively) when adopting the maximum likelihood
approach. Consideration of such an extended family of templates would
necessarily lead to an increased uncertainty in the determination of
parameters and residual correlations due to the MCMC approach. This somewhat
pessimistic conclusion is supported by the results shown in
Table~\ref{tab:alex params} where a change in the SNR of 0.10\% is sufficient
to cause significant changes in the template parameters.

\section{Conclusion}\label{sec:conclusion}

In this work, we have discussed the properties of the noise residuals for
GW150914 based on the maximum likelihood template used
by~\cite{2019JCAP...02..019N}.  We have paid attention to how residual
correlations depend on the template adopted for the description of this event.
In particular, we have considered the ``maximum likelihood’’ template
of~\cite{2019JCAP...02..019N} with extreme spin and the original low-spin
template from~\cite{PhysRevLett.116.061102}.  We have shown that the assertion
in~\cite{2019JCAP...02..019N} that “there are no statistically significant
correlations between the noise residuals of the two detectors at the time of
GW150914” is incorrect.  Specifically, we have demonstrated that the apparent
statistical insignificance of the Hanford/Livingston residual correlations in
the 40\,ms chirp domain is associated with inadequacies in the ML template.
After rematching ML and NR templates to the data, the residual correlations
for these templates converge to 0.50--0.60 in the chirp domain. A study of
various window positions and widths in the provides a more reliable estimation
of residual correlations in the vicinity of the GW150914 chirp and suggests
that the probability that these noise correlations are genuine is roughly
$0.99$ (Section~\ref{sub:more robust} and~\ref{sub:9 window for old
estimator}).

More seriously, the results of Figures~\ref{fig:spin1} and~\ref{fig:spin1b}
suggest that the decision to focus on the single template with maximum
likelihood is flawed in principle.  As noted in Section~\ref{sec:ML brief},
every one of the $37,240$ templates shown in these figures has a high
likelihood. It is natural to expect that the templates of highest likelihood
will be centrally placed among these templates.  Instead, the results reveal a
considerable tension in which the parameters of the ML template are not
consistent with those that characterize the vast majority of the
high-likelihood templates.  In this regard, see also the large changes in the
templates presented in Table~\ref{tab:alex params} that arise from a truly
insignificant $0.1\%$ change in the likelihood.  This situation is
particularly dangerous since the parameters all have names that invite the
reader to attach physical significance to their numerical values.  Our
physical understanding of GW150914 depends critically on whether the black
holes presumed to be involved have the low spins suggested
in~\cite{PhysRevLett.116.061102} or the ultrarelativistic spins
of~\cite{2019JCAP...02..019N}.  The difficulty lies in the fact that the
morphology of the waveforms associated with BBH mergers are remarkably
insensitive to the choice of specific parameter values.  The strength and
importance of this near-degeneracy was considered previously in some detail
in~\cite{Degeneracy} but is largely ignored in most discussions of LIGO’s
results.   Unfortunately, this situation is further complicated by the
presence of six adjustable extrinsic parameters that have nothing to do with
the intrinsic properties of a BBH system\footnote{In this regard, note from
Figure~\ref{fig:spin1} that the right ascension angle, ra, is not well
determined.  As a consequence the sky location of GW150914 cannot be specified
with useful accuracy.}.  The appropriate response to these concerns should be
a careful analysis of the uncertainties associated with the various intrinsic
and extrinsic parameters contained in the waveform.  Unfortunately, no
suitable estimate of parameter uncertainties is provided
in~\cite{2019JCAP...02..019N} or elsewhere.

Finally, we wish to stress that the present manuscript should not be regarded
as either an endorsement of or a challenge to the interpretation of GW150914
as a BBH merger or any other manifestation of gravitational waves.  In our
view, the physical interpretation of this event remains open.  In this sense,
we remain convinced that data must be analyzed and a best common signal
determined without a priori biases and preconceptions before theoretical
models are invoked.  It is a truism that, if gravitational waves are all you
look for, gravitational waves are all you will ever find.

\section{Acknowledgement.}

We would like to thank the authors of~\cite{2019JCAP...02..019N}, Duncan A.
Brown, Collin D.\ Capano, Alex B.\ Nielsen, and Alexander H.\ Nitz, for their
willingness to exchange computer programs and data files.  We also thank James
Patrick Creswell and Sebastian von Hausegger for helpful and stimulating
discussions.  This work has made use of the LIGO software package and data.
Our research was funded in part by the Danish National Research Foundation
(DNRF) and by Villum Fonden through the Deep Space project. Hao Liu is
supported by the Youth Innovation Promotion Association, CAS.

\appendix

\section{The Pearson cross-correlator for LIGO noise}\label{app:correlator}

The cross correlation coefficient between two data sets, $x(t)$ and $y(t)$, at
time $t$ with a time delay $\tau$ and window width $w$ is given by
\begin{equation}
C(t,\tau,w,t) = {\rm Corr}(x_{t+\tau}^{t+\tau+w},y_t^{t+w}).
\label{eqcc}
\end{equation}
Here, ${\rm Corr}(x,y)$ is the Pearson cross-correlation
coefficient~\cite{1895RSPS...58..240P} between two records $x$ and $y$ defined
as
\begin{equation}
{\rm Corr}(x,y) =
\frac{\sum{(x - \overline{x})(y - \overline{y})}} {\sqrt{\sum{(x -
\overline{x})^2} \cdot \sum{(y - \overline{y})^2}}},
\label{corr}
\end{equation}
where the sums extend over all entries contained in the time interval, $w$,
considered and where $\overline{x}$ and $\overline{y}$ are the corresponding
average values of the entries in $x$ and $y$, respectively.

\section{A toy model of residuals}\label{app:toy model}

In this section we present a toy model of how the non-stationarity of the
noise residuals can affect the significance of the amplitude of $C(\tau)$
relative to the importance of the characteristic time lag $\tau$ for their
correlations.

Suppose that the strain data for Hanford and Livingston are modeled as
\begin{eqnarray}
&&S_H(t)=W_H(t)\circledast G(t)+n_H(t), \nonumber\\
&&S_L(t)=W_L(t)\circledast G(t+\tau)+n_L(t) \ ,
\label{a1}
\end{eqnarray}
where $G(t)$ is the true signal, $W$ is a projection operator (including a
transfer function), the operator $\circledast$ denotes a convolution, and
$n(t)$ is noise for Hanford and Livingston. Suppose that we use a template,
$h(t)$, to fit the observational data of eq.~(\ref{a1}).  In general, this
fitting means that
\begin{eqnarray}
&&S_H(t)=W_H\circledast h(t)+W_H(t)\circledast \left[G(t)-h(t)\right]+n_H(t), \nonumber\\
&&S_L(t)=W_L(t)\circledast h(t)+W_L(t)\circledast \left[G(t+\tau)-h(t+\tau)\right]+n_L(t) \, .
\label{a2}
\end{eqnarray}
Here, the noise residuals are given as
\begin{eqnarray}
&&R_H(t)=W_H(t)\circledast \left[G(t)-h(t)\right]+n_H(t), \nonumber\\
&&R_L(t)=W_L(t)\circledast \left[G(t+\tau)-h(t+\tau)\right]+n_L(t)\, .
\label{a22}
\end{eqnarray}
Thus, the cross-correlation between the residuals, $R_H$ and $R_L$, from
Eq.~(\ref{corr}) contains the chance correlations for the Hanford and Livingston
noise and the residuals between the true signal and the template. Note that
the last term is already correlated with the time lag $\tau$. This model
clearly illustrates the non-stationarity and non-Gaussianity of the residuals
for a time domain in the vicinity of the signal even if the noise terms, $n_H$
and $n_L$, are Gaussian and stationary.

In order to investigate the properties of the residual correlations,
$C(\tau)$, for this model, we use the 4096\,s, bandpassed ($35\le
f\le350$\,Hz) and notched strain data given
at~\cite{NBI:Gravitational.waves}, which are identical to those adopted
in~\cite{2019JCAP...02..019N}.  We consider 50 disjoint domains of length
0.2\,s for times well after the GW150914 event and inject in each of them the
signal $G_{\rm losc}$ with an H/L time lag of $\tau=7$\,ms as given by the
LOSC template for GW150914 with masses $m_1=41.743 M_{\odot}$ and
$m_2=29.237M_{\odot}$ and spins, $spin_1=0.355$ and $spin_2=-0.769$. These 50
domains mimic a statistical ensemble with a signal of $G_{\rm losc}$ and
different realizations of genuine LIGO noise.

For the reconstruction of the signal in each of these 50 domains we use the ML
template rematched to the simulated data. It is useful to characterize the
mismatch of the ``signal'' and the template by functions $r_{\rm H/L}$, for
Hanford and Livingston, respectively:
\begin{eqnarray}
r=-f\left(h_{ML}(t)-h_{\rm losc}(t)\right)
\label{a3}
\end{eqnarray}
where the parameter $f$ is chosen such that $|r|$ is much less than  $|h_{\rm
losc}|$ and $|h_{ML}|$. In this toy model all statistical properties of the
noise and the residuals are known. For $f=0$ the $C(\tau)$ from eq.~(\ref{corr})
corresponds to the pure noise correlations, for $f=1$ we get the contribution
from the difference of the templates plus noise and, formally, for
$f\rightarrow\infty$ we should see that $C\rightarrow -1$ for $\tau=7$\,ms in
each domain.

\begin{figure}[!tbh]
\centerline{
\includegraphics[width=0.48\textwidth]{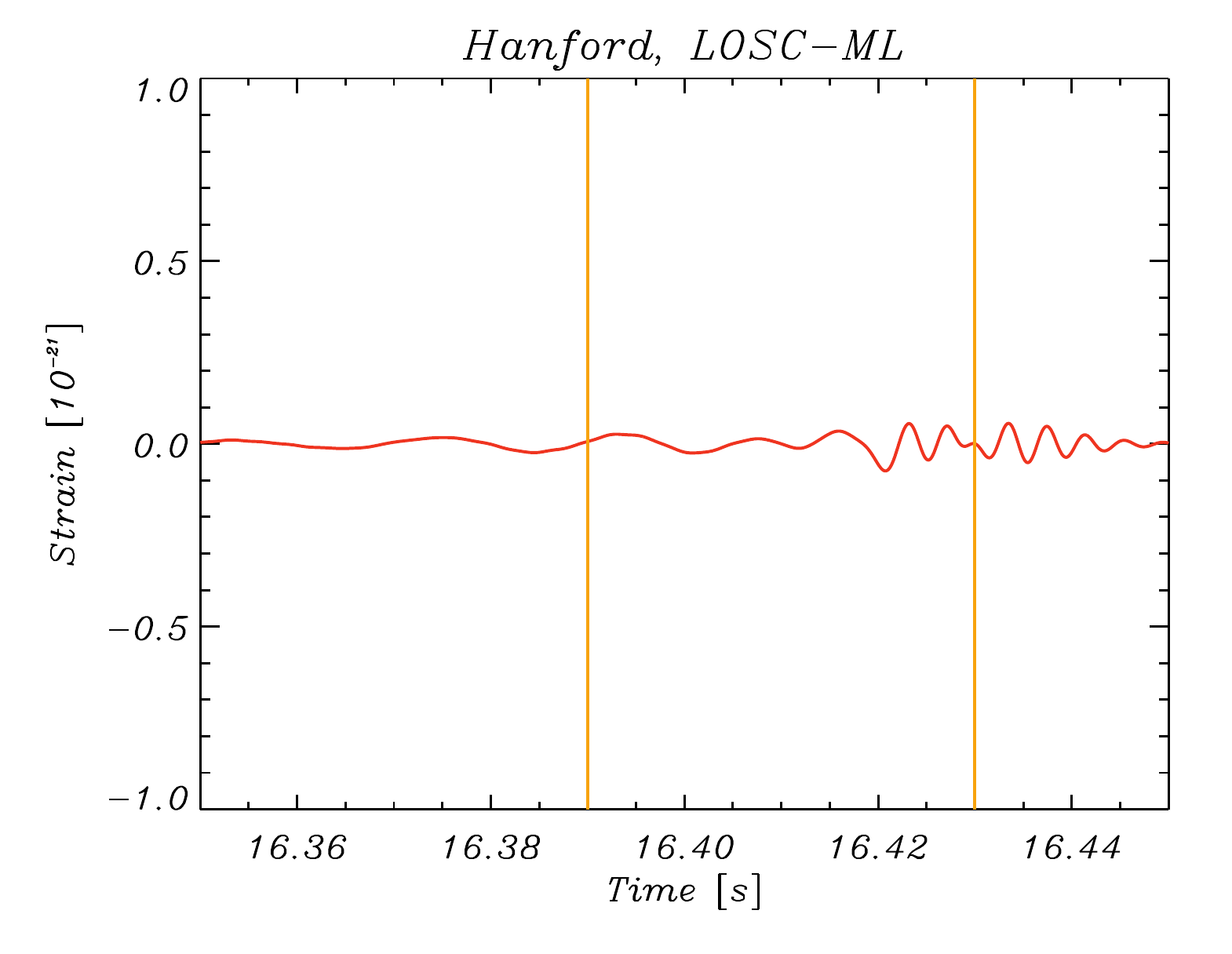}
\includegraphics[width=0.48\textwidth]{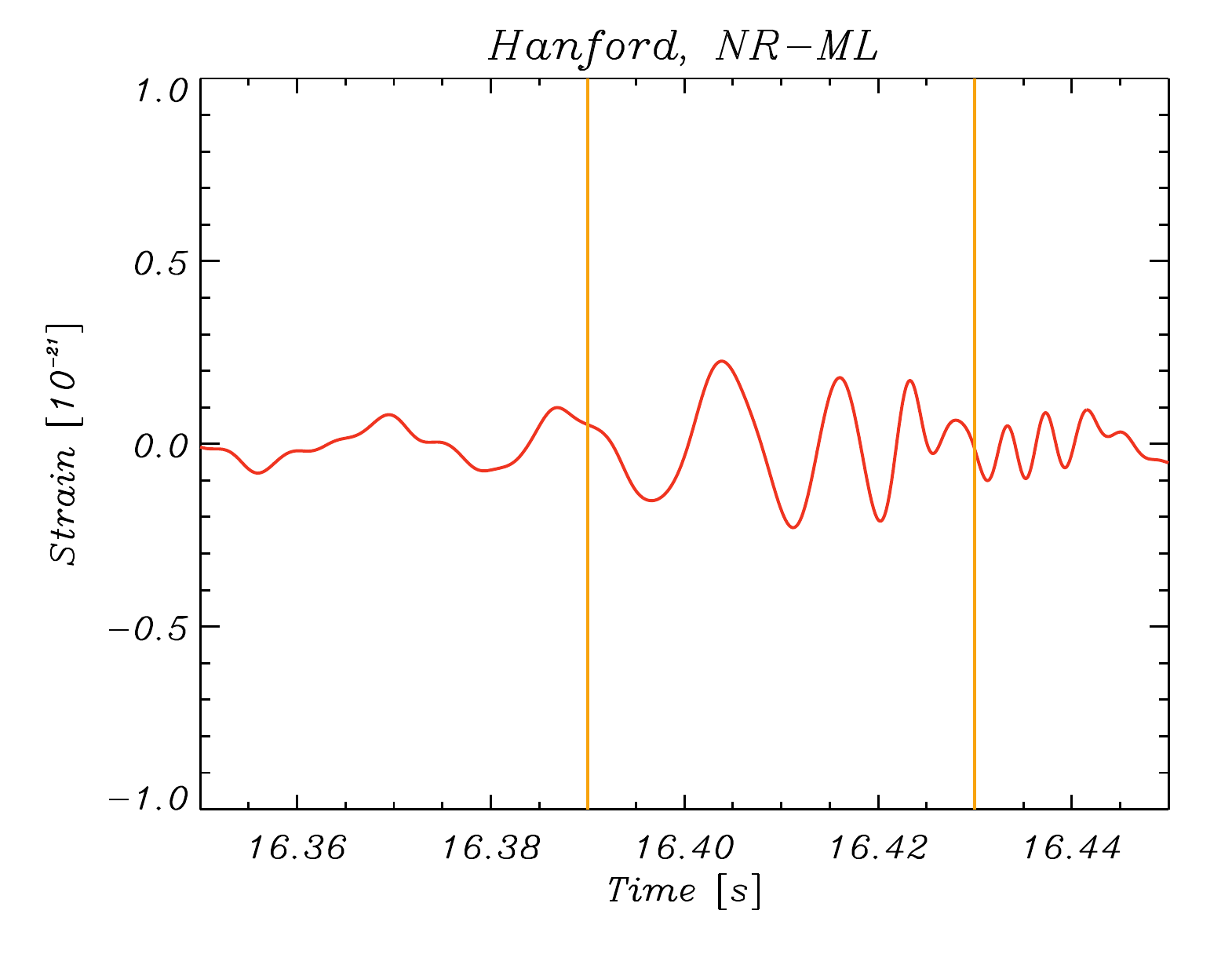}
}
\caption{ Left panel. The difference between ML and LOSC templates. Right
panel. Same as the left but for ML and NR templates }\label{fig:app:f1}
\end{figure}

Figure~\ref{fig:app:f1} shows the difference between the (bandpassed and
notched) LOSC and ML templates.  These templates are matched to the ML
template used as a proxy for the GW150914 signal.
For comparison, the figure also shows the difference between NR and ML
templates with the same normalisation. The difference between NR and ML
templates is seen to be roughly twice the size of the difference between the
ML and LOSC templates. Note the position of the peaks is different in the two
panels of this figure.

\begin{figure}[!tbh]
\centerline{
\includegraphics[width=0.48\textwidth]{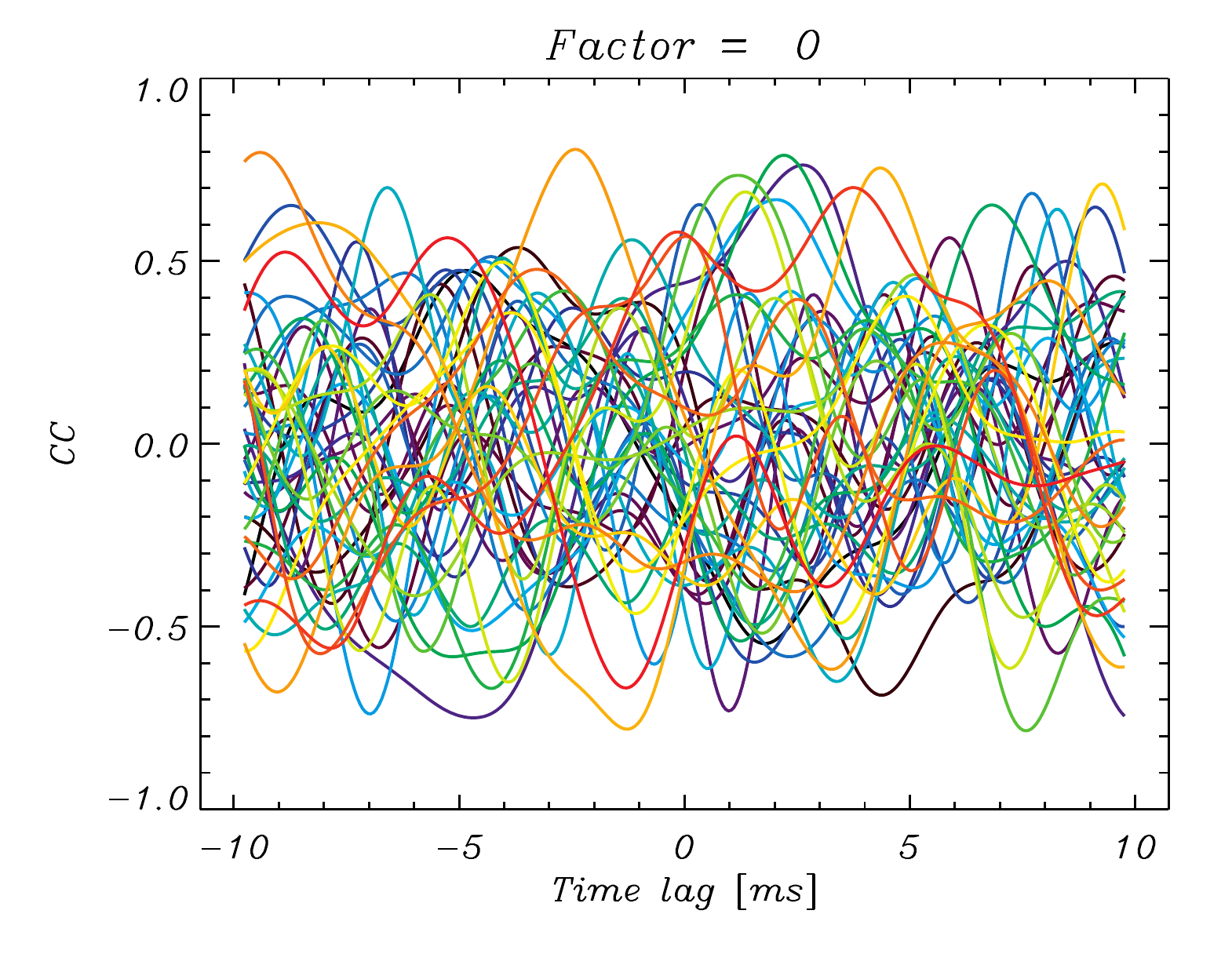}
\includegraphics[width=0.48\textwidth]{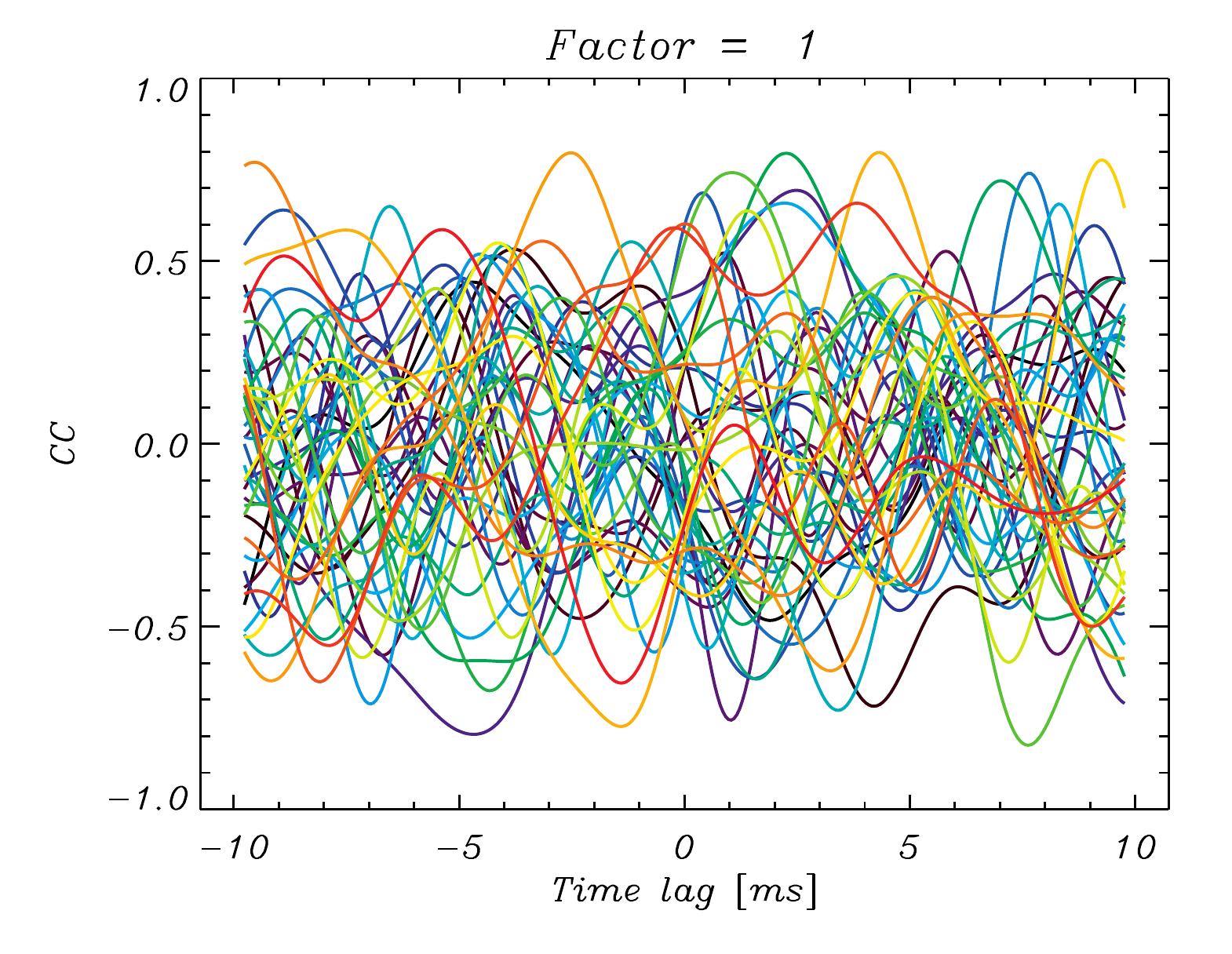}
}
\caption{ The residual correlation, $C(\tau)$ for $f=0$ (left panel) and $f=1$
(right panel). }
\label{f2}
\end{figure}
In Figure\ref{f2} the left panel corresponds to the pure chance correlations
of the noise and the right panel show the residual correlations for the choice
$f=1$.  Note that the choice $f=1$ corresponds to the case for which the
signal (in the form of the LOSC template) is fitted by the ML template. Let us
focus on the time lag of the residual correlations in the vicinity of $\tau =
7$\,ms. For $f=0$, one sees a negative peak in this domain with $C\simeq
-0.75$. (See the green line in Figure\ref{f2}). According to Figure 8, the
corresponding probability for finding a peak of this size is $P\simeq 0.01$.
For the choice $f=1$, the amplitude of this peak becomes $C\simeq -0.9$ a
corresponding probability of $P\simeq 10^{-4}$ according to Figure 8.  The
same tendency can be seen for the yellow positive peak at $\tau\simeq
-(3.5)$\,ms. For $f=0$ the chance noise correlations are $C(\tau\simeq
-3.5)\simeq 0.8$ with a corresponding probability of $P\simeq 10^{-3}$ which
decreases to $P\simeq 2 \times 10^{-4}$ for $f=1$.

Thus, in the toy model considered here, we can trace the influence of the
mismatch of signal and template on the probability of the peak correlations.
Usually, a mismatch between template and signal leads to a stronger residual
(anti)correlation than that found for chance noise correlations alone.
However, this is not necessarily the case when the genuine signal is unknown,
as is the case in the chirp domain of the GW150914 event.

\subsection{Residuals correlations for the high amplitude case.} 

\begin{figure}[!tbh]
\centerline{
\includegraphics[width=0.32\textwidth]{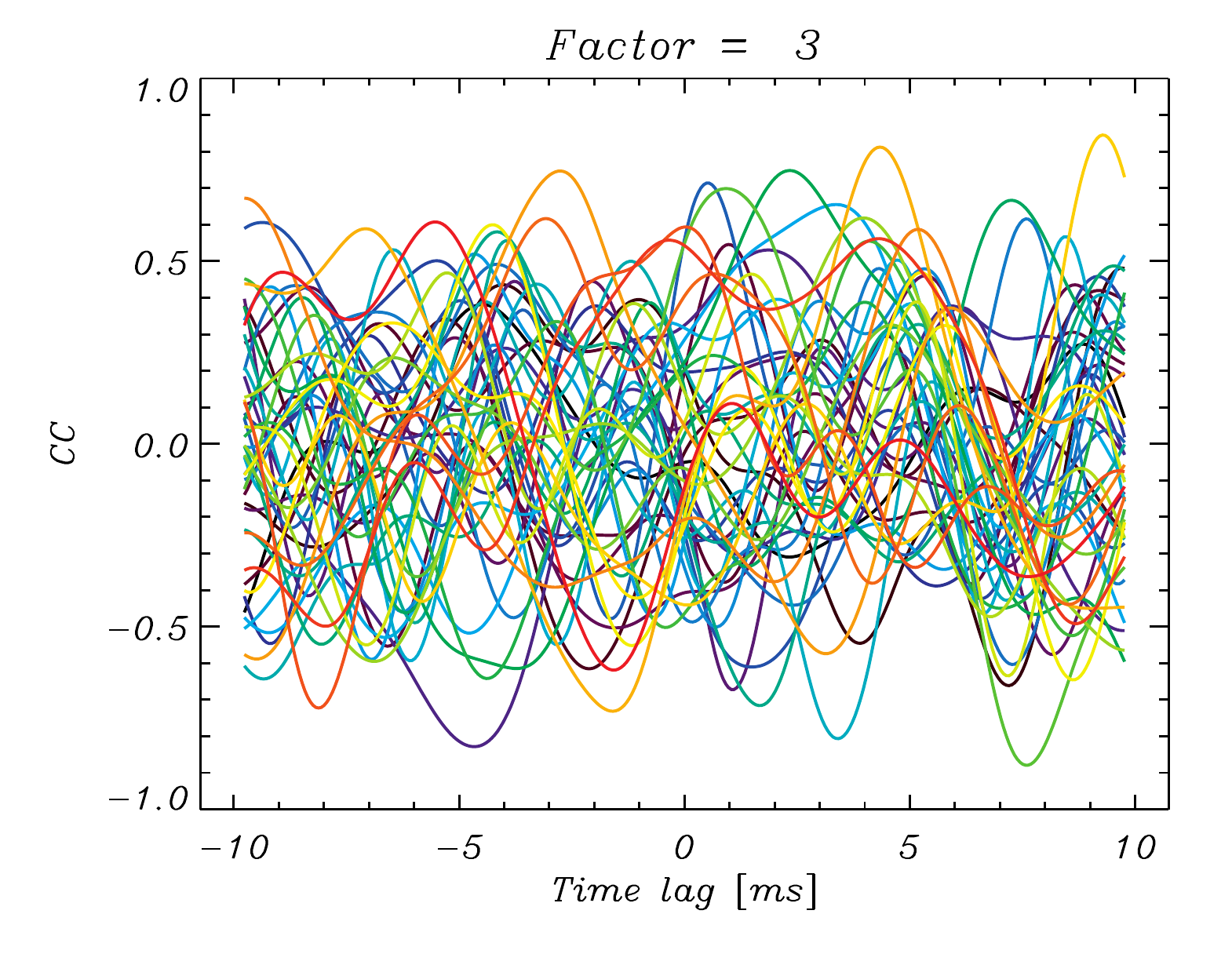}
\includegraphics[width=0.32\textwidth]{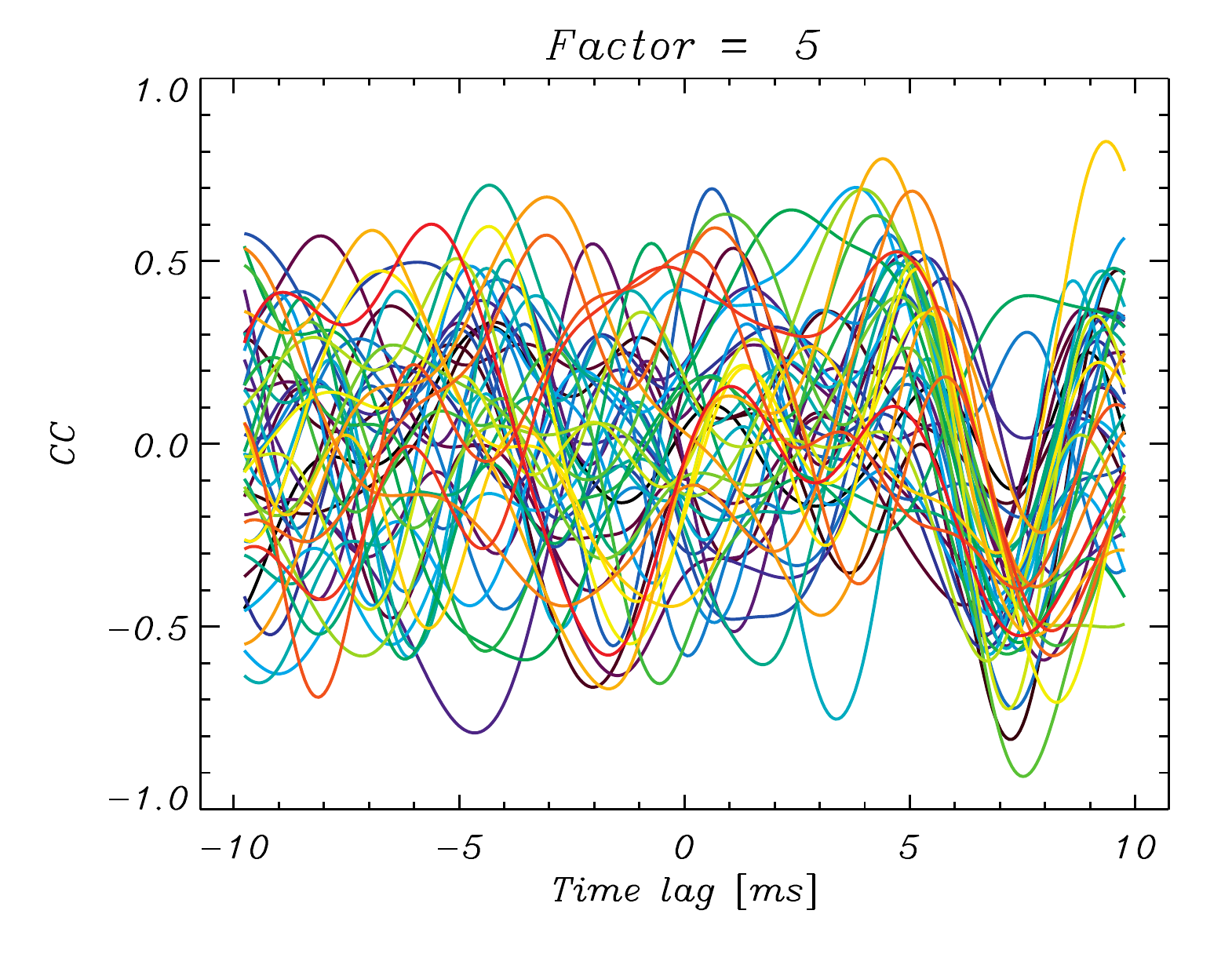}
\includegraphics[width=0.32\textwidth]{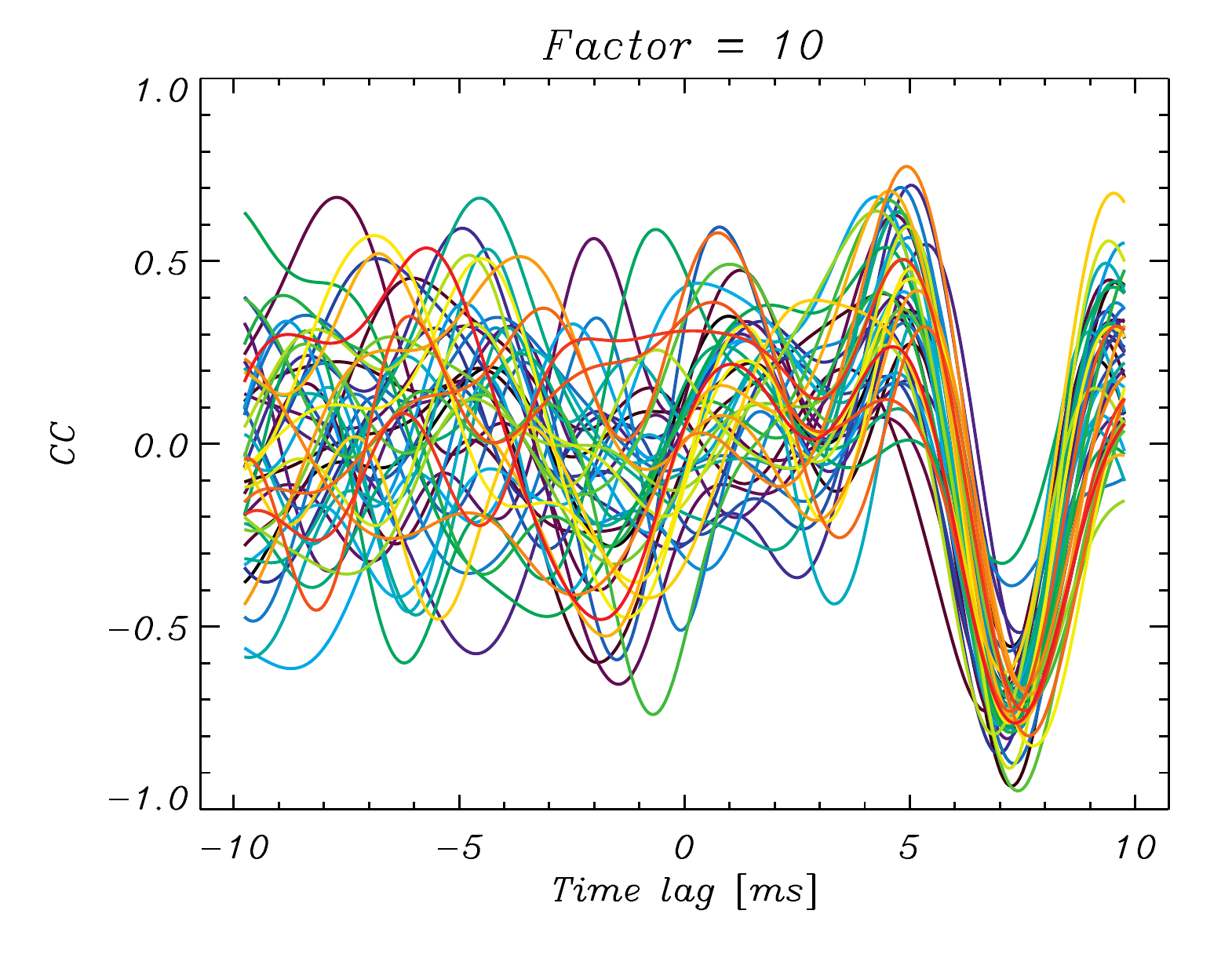}
}
\caption{The residual correlation $C(\tau)$ for $f=3$ (left panel), $f=5$
(middle panel) and $f=10$ mode (right panel). }
\label{f3}
\end{figure}

In the toy model presented above the true signal is given by the LOSC template
is approximated by the ML template. In this case, the ``exact'' solution to
the problem of residual correlations is given by eq.~(\ref{a2}) with $f=1$. Let
us now assume that we have no information regarding chance correlations, so
that we attempt to fit the data using a combination of the ML template and the
residuals, $r$ from eq.~(\ref{a3}), with $f>1$.  Note that the residuals $r$
have a characteristic time lag $\tau\simeq 7$\,ms. Figure~\ref{f3} shows
$C(\tau)$ for $f=3,5$, and $10$ as a function of $|\tau|\le 10$\,ms.
 
We see that, as the residuals, $r$ become larger, almost all of the
cross-correlation functions collapse to asymptotic values, $C(\tau)$, with
clearly established time lag 7\,ms,
\begin{figure}[!tbh]
\begin{center}
\includegraphics[width=0.48\textwidth]{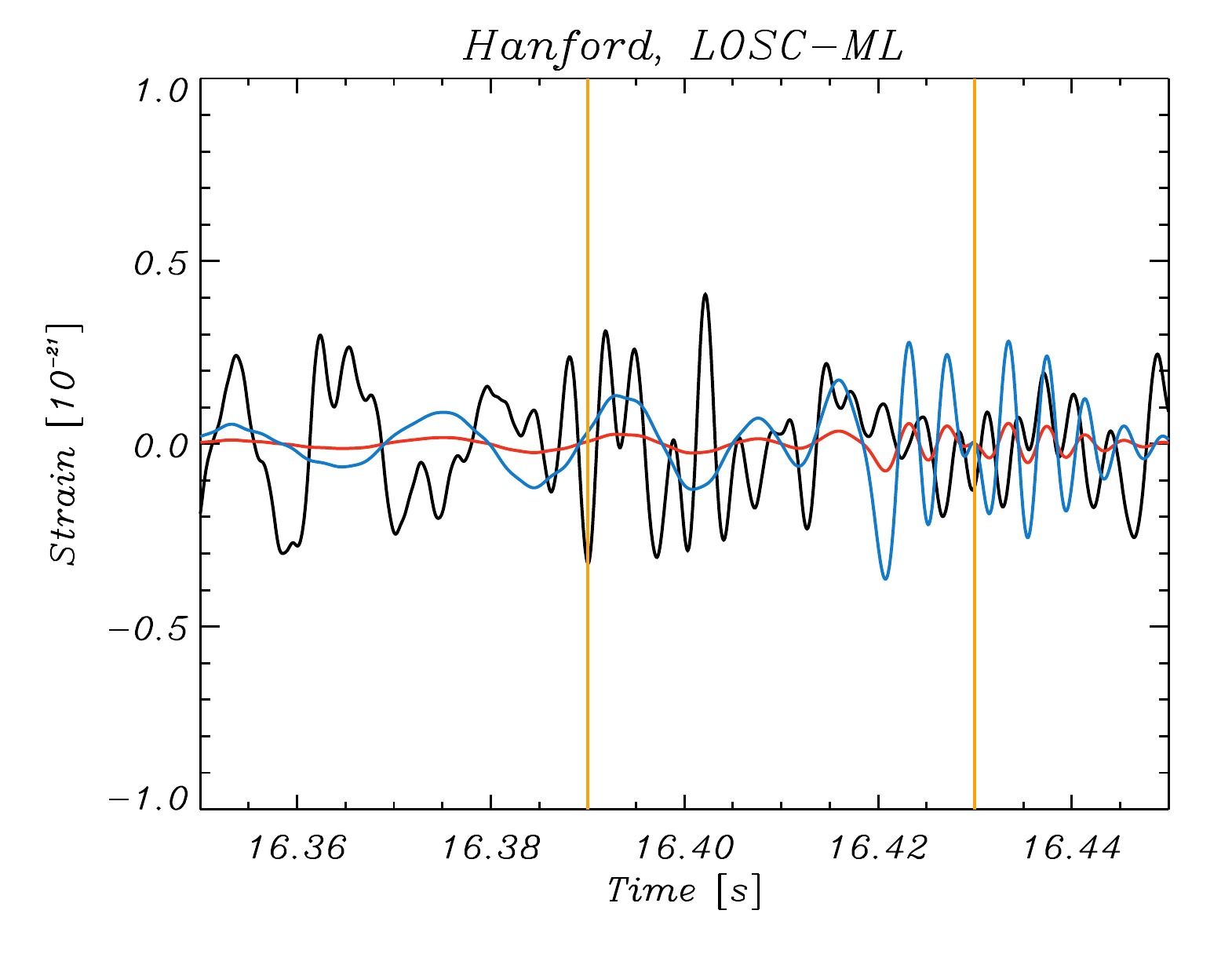}
\includegraphics[width=0.48\textwidth]{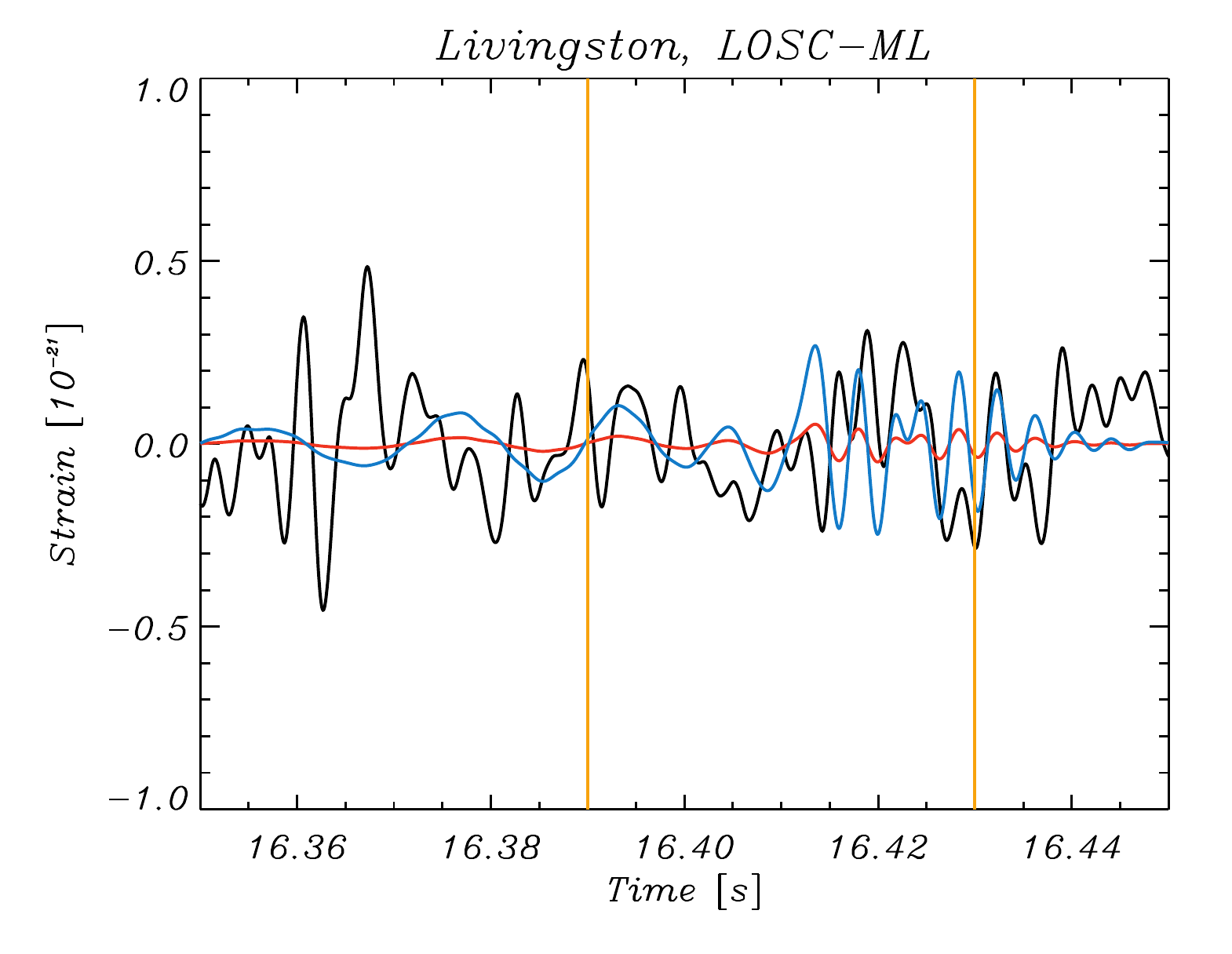}
\end{center}
\caption{ Comparison of the Hanford (left) and Livingston (right) ML template
residuals (black) with the $f\left[h_{losc}(t)-h_{ML}(t)\right]$ term, where
$f=1$ and 5 respectively (red and blue). Two yellow vertical lines indicate
the position of chirp domain. }
\label{f4}
\end{figure}
Moreover, the comparison between $f=5$ and $f=10$ models presented in
Figure~\ref{f4}, clearly illustrates the transition of the chance correlations
from positive to negative values in the domain $\tau=7$\,ms (See the blue and
green lines in this domain.) It is worth noting that the term 
$f\left[h_{\rm losc}(t)-h_{ML}(t)\right]$ becomes roughly comparable to the
residuals (black line) when $f=5$. This collapse of correlations is due to the 
increasing amplitude of the residuals relative to the amplitude of the noise
that makes the residual term more non-stationary and non-Gaussian. At the same
time, the application of the Gaussian statistics from the right panel of
Figure~\ref{fig:pos ABC} will give us a completely misleading result. It
follows from the right panel of Figure~\ref{fig:pos ABC} that the
corresponding chance probability is localized in the domain $0.1-1$.

It is worth noting that the effect of the collapse of the residual
correlations, described with our toy model, is very general. For unknown
signal, $g(t)$, the noise residuals are given by eq.~(\ref{a3}), where both $g$
and $h$ have the same time lag $\tau$. Thus, when the term $\propto (g-h)$ is
close to the noise term $n(t)$, we will see the beginning of the collapse of
$C(\tau)$ around $\tau\simeq 7$\,ms. The residuals are non-stationary and
non-Gaussian, but the amplitude of $C(\tau)$ can be relatively small
($C(\tau\simeq 7ms)\simeq 0-0.5)$. Once again, assuming Gaussianity of the
residuals and using Figure~\ref{fig:pos ABC} one will conclude that this
particular realisation of the residuals is very likely, which is obviously not
the case.


\providecommand{\href}[2]{#2}\begingroup\raggedright\endgroup

\end{document}